\documentclass[12pt]{article}
\usepackage[top=1in, bottom=1in, left=1in, right=1in]{geometry}
\pdfoutput=1

\usepackage{natbib}
 \bibpunct{(}{)}{;}{a}{}{,} 
\usepackage{graphicx}            
\usepackage[cmex10]{amsmath}
\usepackage{amssymb}
\usepackage{amsthm}              
\usepackage{mathtools}
\usepackage[pdftex,colorlinks=true,linkcolor=blue,citecolor=blue,urlcolor=blue,bookmarks=false,pdfpagemode=None]{hyperref}
\usepackage{url}
\usepackage{verbatim}
\usepackage{fancyhdr}
\usepackage{setspace}
\usepackage{paralist}
\usepackage{lineno}

\usepackage{tikz}
\usepackage{enumerate}
\usepackage{bm}
\usepackage{dsfont} 

\newcommand{\Ni}{\mathcal{N}}
\newcommand{\Yzero}{\textbf{Y(0)}}

\newcommand{\Yw}{Y(Z)}
\newcommand{\W}{Z}
\newcommand{\Y}{Y}

\newcommand{\svar}{\mathrm{var}}
\newcommand{\sexp}{E}
\newcommand{\scov}{cov}
\newcommand{\sexpy}{\mathbb{E}_\Theta}
\newcommand{\svary}{\mathbb{V}_\Theta}
\newcommand{\scovy}{\mathbb{C}\text{ov}_\Theta}
\newcommand{\sexpz}{\mathbb{E}_{\mathcal{Z}}}
\newcommand{\svarz}{\mathbb{V}_{\mathcal{Z}}}
\newcommand{\scovz}{\mathbb{C}\text{ov}}
\newcommand{\zb}{\mathcal{Z}^b}
\newcommand{\zbo}{\mathcal{Z}^{b} \cap \mathcal{Z}^o}
\newcommand{\zbu}{\mathcal{Z}^{b} \cap \mathcal{Z}^u}
\newcommand{\zbuo}{\mathcal{Z}^{b} \cap \mathcal{Z}^u \cap \mathcal{Z}^o}

\newcommand{\mse}{\textsc{mse}}
\newcommand{\MSE}{mean square error}

\newcommand{\Scal}{\mathcal{S}}
\newcommand{\bias}{\text{Bias}}
\newcommand{\iv}{\mathds{1}}

\newtheorem{remark}{Remark}
\newtheorem*{constraint}{Regularity conditions}

\newtheorem{theorem}{Theorem}
\newtheorem{example}{Example}
\newtheorem{lemma}{Lemma}
\newtheorem{definition}[theorem]{Definition}
\newtheorem{corollary}[theorem]{Corollary}

\begin{document}
\pagestyle{empty}

\title{
Model-assisted design of experiments in the presence of network correlated outcomes
\protect\thanks{Guillaume W. Basse is a graduate student in the Department of Statistics at Harvard University (\href{mailto:gbasse@fas.harvard.edu}{gbasse@fas.harvard.edu}). Edoardo M.~Airoldi is an Associate Professor of Statistics at Harvard University (\href{mailto:airoldi@fas.harvard.edu}{airoldi@fas.harvard.edu}).
This work was partially supported 
 by the National Science Foundation under grants 
  CAREER IIS-1149662 and IIS-1409177,
 and by the Office of Naval Research under grant 
  YIP N00014-14-1-0485. 
 Guillaume W. Basse is a Google Fellow in Statistics.
 Edoardo M.~Airoldi is an Alfred Sloan Research Fellow, and a Shutzer Fellow at the Radcliffe Institute for Advanced Studies.
The authors are grateful to Iavor Bojinov and Donald B.\ Rubin for  constructive discussion and comments.}}

\author{
 Guillaume W. Basse, Edoardo M. Airoldi\\
 Department of Statistics\\
 Harvard University, Cambridge, MA 02138, USA}
\date{}

\maketitle
\thispagestyle{empty}
\newpage


\begin{abstract}
We consider the problem of how to assign treatment in a randomized experiment, in which the correlation among the outcomes is informed by a network available pre-intervention. Working within the potential outcome causal framework, we develop a class of models that posit such a correlation structure among the outcomes. Then we leverage these models to develop restricted randomization strategies for allocating treatment optimally, by minimizing the  mean square error of the estimated average treatment effect. Analytical decompositions of the mean square error, due both to the model and to the randomization distribution, provide insights into aspects of the optimal designs. In particular, the analysis suggests new notions of balance based on specific network quantities, in addition to classical covariate balance. The resulting balanced, optimal restricted randomization strategies are still design unbiased, in situations where the model used to derive them does not hold. We illustrate how the proposed treatment allocation strategies improve on allocations that ignore the network structure, with extensive simulations.\\
\vfill

\end{abstract}


\newpage
\tableofcontents


\newpage
\section{Introduction}

The past decade has witnessed a surge of interest in causal analyses in the context of social networks, social media platforms and online advertising \citep[e.g.,][]{christakis2007spread, aral2009distinguishing, bakshy2011everyone, bakshy2012role, bond201261, kim2015aa, gui2015aa, phan2015aa, cavusoglu2016aa}. 
From a statistical perspective, the challenging aspect of these applications is how to account for the presence of connections, or network data, observed pre-intervention, possibly with uncertainty, and often missing \citep{airoldi2015aa}.

While there is a well-developed literature on several aspects of the statistical analysis of network data \citep[e.g.,][]{wasserman1994social, kolaczyk2009statistical,goldenberg2010survey,bickel2009nonparametric}, 
the literature about methods for experimentation and causal analyses that leverages observed connections is at a budding stage \citep[e.g.,][]{rosenbaum2007aa, hudgens2008toward,aronow2013estimating,toulis2013aa,ogburn2014,sussman2016aa}. 

Moreover, even when considering the average treatment effect as the inferential target of interest, multiple conceptualizations and assumptions are possible, which require different methodological approaches to achieve valid inference when analyzing the same experiment \citep{karwa2016aa,sussman2016aa,airoldi2015ab}.

Here, we consider the problem of how to  assign treatment in a randomized experiment, when the correlation among the outcomes is informed by a network available at the design stage.


\subsection{Related work}

The need to account for network connections in causal analyses has led scholars to focus on two specific problem settings: 
 (i) network interference \citep{airoldi2012, toulis2013aa, aronow2013estimating, ugander2013graph, eckles2014design}, a situation where the potential outcomes of unit $i$ are a function of the treatment assigned to unit $i$ and of the treatment assigned to other units that are related to unit $i$ through the network, or  of the observed outcomes of these related units,
 (ii) network-correlated outcomes \citep{mcpherson2001birds,manski2013identification,shalizi2011homophily}, 
 an alternative setting where the network informs the correlation among the potential outcomes, because the potential outcomes of unit $i$ are a function of its covariates and those of other units, and the covariates of units that are connected are more similar than the covariates of the units that  are not. In this paper, we focus on the network-correlated outcomes setting which has received less attention.
With few exceptions \citep{aral2009distinguishing,shalizi2011homophily}, the literature considers these problems in isolation even as it is motivated by scenarios in which both are plausible \citep[e.g., see][]{christakis2007spread}. 
The effects of network interference, whether as the target of inference or as a nuisance, have been mostly studied in randomized experiments \citep{parker2011, airoldi2012, toulis2013aa, aronow2013estimating, ugander2013graph, eckles2014design, karwa2016aa, sussman2016aa} with a recent exception \citep{forastiere2016aa}. 

The confounding due to correlated outcomes, typical of problems where homophily is plausible \citep{mcpherson2001birds}, has been mostly explored in observational studies using potential outcomes \citep{manski2013identification} or causal graphical models \citep[e.g., see][]{shalizi2011homophily, shalizi2016aa}.
\citet{aral2009distinguishing} proposed a randomization strategy to disentangle interference and homphily in an application to online marketing, in the context of a dynamic network. However, this randomization strategy is tailored to the application and hard to analyze theoretically.

We have been working toward an analytical understanding of how to best identify and estimate the causal effect of interference in the presence of confounding due to correlated outcomes, in the context of randomized experiments on networks. Our approach is to  develop restricted randomization strategies that leverage a (static) network available pre-intervention. 
%
Thus far, we  analyzed shortcomings of popular randomization strategies for estimating causal effects  \citep{karwa2016aa}, and developed elements of a theory of estimation for them \citep{airoldi2015ab,sussman2016aa}, in the presence of network interference. We leveraged these results to estimate causal effects in observational studies \citep{forastiere2016aa}. This body of work has already led to some insights and general principles \citep{airoldi2015aa}.


In this paper, we propose a collection of model-assisted restricted randomization strategies to improve estimation of the average treatment effect, in the experiments where a network is available pre-intervention, in the presence of network-correlated outcomes.
Restricted randomization as a way to increasing the precision of estimates has a long tradition in the field of experimental design \citep[e.g., see][]{yates1948aa,youden1972aa,simon1979aa,bailey1983aa,higham2015aa}. The basic idea is that some 
assignments are considered problematic (e.g., when some covariates are unbalanced between the treatment arms) and can be excluded. In networks,
the challenge is to identify which are the features that must be balanced, which makes it difficult to know how to restrict the randomization.

\subsection{Contributions}

Our approach is inspired by classical work on model-assisted survey sampling in which a specific model is used to reduce the variance of a given estimator (typically, a liner weighted estimator) in a way that does not harm its properties when the model is wrong \citep[e.g., see][]{sarndal2003model}. 

Drawing inspiration from the model-assisted survey sampling literature, we propose a model-assisted approach to the design of experiments.
Namely, 
we posit a working model for the potential outcomes specified conditionally on a network observed pre-intervention, and then restrict the randomization to 
assignments for which the estimator of interests achieves a low mean square error. The class of models we propose leads to analytical expressions for the 
mean square error that suggest new notions of balance in terms of network statistics related to the degree distribution. We also develop new theoretical 
results showing that the model-assisted restricted randomization approach we propose maintains the design unbiasedness of the difference-in-means estimator
even when the model is misspecified, and reduces the expected variance of the estimator when the model holds.

\section{Analytical insights for evaluating allocations} 
\label{sec:nsm-analysis}


Here we introduce the model-assisted approach to designing experiments; we posit a model for the potential outcomes that is used for calculating the mean square error of the difference of means estimator.
Explicit formulas for the mean square error, for fixed allocation vector $Z$ and averaged over allocation vectors $Z$s, indicate which aspects of the network play a role in the estimation of causal effects, in this setting, in Sections \ref{sec:cond-mse} and \ref{sec:marg-mse}.
We then introduce a more general model in Section \ref{sec:generalization} and we show, in the appendix, the extent to which the intuition developed in the illustrative model holds more generally. 
While the models we posit help developing analytical insights, and are useful in specific applications,  the methodology we propose is design-unbiased even when these models do not hold, as we show in Section~\ref{sec:theory}.


\subsection{Causal inference setup}


We work within the potential outcomes framework \citep{rubin1974aa,holland1986,imbens2015aa}.
We consider a population of $N$ units, a binary treatment, denoted $Z_i = 1$ if unit $i$ is assigned to treatment, and real-valued outcomes, denoted $Y_i$. The corresponding vectors are denoted $\Y,\W$, respectively. We assume the stable unit-treatment-value assumption holds, which implies that the outcome of unit $i$ is only a function of the treatment assigned to it, $Y_i(\W)=Y_i(Z_i)$, thus excluding interference \citep{rubin1974aa}.


We consider a finite population setting, where the potential outcomes $\Yw$ are unknown constant quantities, given $\W$.
The only source of variation is how treatment is allocated to units. 
We assume treatment is allocated according to a distribution 
on the space of all binary vectors of length $N$, typically referred to as the randomization distribution \citep{imbens2015aa}.


For illustrating the idea of model-assisted restricted randomizations, we consider the average treatment effect as the the inferential target of interest, defined as $\tau^* = (1 / N) \sum_{i=1}^N \{ Y_i(1) - Y_i(0) \}$, and the popular difference-in-means estimator for the average treatment effect,
\begin{equation}\label{eq:naive-est}
	\hat{\tau}(\Y \mid \W^{}) = \textstyle \frac{\sum_{i=1}^N Z_i Y_i}{\sum_{i=1}^N Z_i} - \frac{\sum_{i=1}^N (1-Z_i) Y_i}{\sum_{i=1}^N (1-Z_i)}.
\end{equation}


\subsection{The normal-sum model}
\label{sec:normal-sum}

The model-assisted approach to experimental design requires a model, which is used to improve the inferential properties of the difference-in-means estimator when the model holds.
We posit a the model that depends on a network, which is available at the design stage in our setup.

Consider $N$ units and an undirected network $\mathcal{G}$ among them, or, equivalently, a binary adjacency matrix $A$ of size $N\times N$, with the added constraint that $A_{ii} = 1$ for all $i$, which we refer to as the extended adjacency matrix. The neighborhood of a unit $i$ is defined as the index set $\mathcal{N}_i = \{ j \hbox{ s.t. } A_{ij}=1 \hbox{ or } A_{ji}=1 \}$. 
Let us posit the following model,
\begin{flalign}
 \label{eq:nsx}
  X_j &\overset{iid}{\sim} \hbox{ Normal }(\mu, \sigma^2) \\
 \label{eq:nsy0}
  Y_i(0) \mid X&\overset{ind}{\sim} \hbox{ Normal }(\textstyle \sum_{j \in \Ni_i} X_j, \gamma^2) \\
 \label{eq:nsy1}
  Y_i(1) &= Y_i(0) + \tau.
\end{flalign}
The  network induces correlation among the outcomes that are assigned to control because the mean of each $Y_i(0)$ is given by the sum of the covariate values, $X_j$, of units $j$ in a neighborhood of unit $i$. The effect of treatment is additive.
Equations \eqref{eq:nsx}--\eqref{eq:nsy1} define the normal-sum model.
The implied model for the observed outcomes, $\Y^{\rm obs}$, is given in Appendix \ref{sec:yobs}. We consider a more general version of this
model in Section~\ref{sec:generalization}, but will otherwise focus on this model for the sake of clarity in presenting the restricted randomization approach.

For our purposes, the normal-sum model provides a useful abstraction for exploring the problem of optimal design of experiments in the presence of network-correlated outcomes. However, an illustrative application will help anchor the intuition. 
The normal-sum model arises naturally, for example, when considering the time users spend on a social media platform. 
Consider the binary treatment $Z_i$ to be the exposure to a new feature of the website designed 
to increase engagement and time spent online, and let $Y_i(Z_i)$ be the time spent online by user $i$  when assigned to treatment $Z_i$. The causal effect of interest $\tau$ is then the effect of the new feature on the time spent online. Let us assume a constant, additive treatment effect, for simplicity. 
In the absence of network connections and in the absence of treatment (i.e., $Z_i=0$), $X_i$ is the expected value of $Y_i(0)$ conditional on $X_i$. So $X_i$ can be thought of as the intrinsic propensity of user $i$ to spend time on the website. The model then captures the fact that the time spent on the website by user $i$ increases with the number of its neighbors,  with the neighbors'  propensities to spend time on the website, and with the exposure to the new feature if the treatment has an effect.

In the appendix, we also consider a variant of the normal-sum model, which we refer to as the normal-mean model, in which the mean of $Y_i(0)$ is given by the average of the covariate values, $X_j$, of units $j$ in a neighborhood of unit $i$.
%
%
In Section \ref{sec:generalization} we consider a general family of models that subsumes the normal-sum and normal-mean models. Model-assisted design strategies for models in this family, similar to those developed in Sections \ref{sec:nsm-analysis} and \ref{sec:tm} for the normal-sum model, can be developed by using the mean-square error calculations analysis detailed in Appendix \ref{section:general-model}.

The rest of the paper explores the implications of the normal-sum model for designing optimal treatment allocation strategies.


\subsection{Interpretation of the mean square error for a fixed treatment allocation vector}
\label{sec:cond-mse}
	
We compute the mean square error of the difference-in-means estimator, according to the normal-sum model for $\Y^{\rm obs}$, defined as $\mse(\hat{\tau} \mid \W) \equiv \sexp\{(\hat{\tau} - \tau^*)^2 \mid \W\}$, for a fixed treatment allocation vector $\W$. We refer to this quantity as the conditional mean square error. 
%
%
We have,
\begin{equation}\label{eq:rerand}
\mse(\hat{\tau} \mid \W) = 
 \underbrace{\mu^2 \{\delta_{\mathcal{N}}(\W)\}^2}_{\text{bias}^2} +  
 \underbrace{\gamma^2 \omega(\W)^T \omega(\W) + \sigma^2 \omega(\W)^T  A^T A \omega(\W)}_{\text{variance}}.
\end{equation}
We can identify desirable assignments by evaluating their conditional mean square error. This idea is the basis for the model-assisted restricted randomization strategies, in Section~\ref{sec:marr}. 

In the absence of specific constraints on the number of treated units, different treatment allocation vectors will generally have a different number of treated and untreated units, defined $N_1 = \sum_i Z_i$ and $N_0=\sum_i (1-Z_i)$ respectively, both functions of $\W$.
Then the bias term is
\begin{equation}\label{interpretation:bias}
 \mu \cdot \delta_{\mathcal{N}}
  ~ = \mu \,\cdot\bigm( \textstyle \frac{1}{N_1} \sum_{\{i:Z_i=1\}} \mid\Ni_i\mid - \frac{1}{N_0} \sum_{\{i:Z_i=0\}} \mid\Ni_i\mid \bigm).
\end{equation}
The bias is proportional to the difference in the average neighborhood sizes of treated and untreated units. Intuitively, this difference measures a lack of balance between the two groups, in terms of their network characteristics---specifically, the average degree. A larger value of the mean $\mu$ amplifies the contribution of this imbalance to the mean square error. 
Since the designer does not have control over $\mu$, desirable treatment assignments minimize bias by balancing the average neighborhood size between treated and untreated units.
The first variance term is
\begin{equation}\label{interpretation:var1}
 \gamma^2 \omega^T \omega
  ~ = \gamma^2 \bigm(\textstyle\frac{1}{N_1} + \frac{1}{N_0}\bigm) ,
\end{equation}
which is minimized when $N_1 = N_0$. Intuitively, this term penalizes the difference between the number of treated and untreated units. A larger value of the parameter $\gamma$ amplifies the contribution of this imbalance to the mean square error. 
This result is consistent with  classical results on the optimality of balanced randomizations for estimating the average treatment effect in the absence of network correlated outcomes \citep{fisher1954methods,imbens2015aa}.
%
%
The second variance term involves features of the network; it is
\begin{flalign}
\label{eq:31} 
\sigma^2 \cdot \omega^T  A^T A \omega 
	 & \textstyle = \frac{\sigma^2}{N_1^2} \cdot \sum_{\{i,j : Z_i = Z_j=1\}} \mid\mathcal{N}_i \cap \mathcal{N}_j\mid \\ 
\label{eq:32} 
	 & \textstyle + \frac{\sigma^2}{N_0^2} \cdot \sum_{\{i,j : Z_i = Z_j=0\}} \mid\mathcal{N}_i \cap \mathcal{N}_j\mid \\ 
\label{eq:33} 
	 & \textstyle - \frac{2\sigma^2}{N_1\cdot N_0} \cdot \sum_{\{i,j : Z_i = 1 \text{ and } Z_j=0\}} \mid\mathcal{N}_i \cap \mathcal{N}_j\mid.
\end{flalign}
%
%
The factor on the right hand side of \eqref{eq:31} is proportional to the average number of shared neighbors among pairs of units both assigned to the treatment group. The factor in \eqref{eq:32} is proportional to the average number of shared neighbors among pairs of units both assigned to the control group. The factor in \eqref{eq:33} is proportional to the average number of shared neighbors among pairs of units, one assigned to treatment and one assigned to control. 
Considering the signs in front of these three factors, the second variance term may be minimized by assigning units with shared neighbors to different groups, and by avoiding assigning treatment or control to entire clusters of units that are densely connected.


\subsection{Interpretation of the mean square error averaged over allocation vectors}
\label{sec:marg-mse}

Next, we compute the mean square error of the difference-in-means estimator, according to the normal-sum model and the distribution on the allocation vectors implied by a complete randomization strategy---which assigns equal probability to all of the treatment allocation vectors $\W$ for which the numbers of units in treatment and control are fixed to $(N_0,N_1)$.
We refer to this quantity, defined as $\mse(\hat{\tau}) \equiv \sexp[\sexp\{(\hat{\tau} - \tau^*)^2 \mid \W\}]$, as the marginal mean square error.
It is, 
\begin{flalign}
 \label{eq:fullrand1}
	\mse(\hat{\tau}) 
	&  \textstyle = \bigm(\frac{1}{N_1} + \frac{1}{N_0}\bigm)  \bigm(\gamma^2 + \sigma^2\bigm) + \\
 \label{eq:fullrand2}
  	& \textstyle + \bigm(\frac{1}{N_1} + \frac{1}{N_0}\bigm) \bigm\{\underbrace{\textstyle \sigma^2 (\overline{\mid\Ni\mid} - 1)}_{C_1}- \underbrace{\textstyle \frac{2 \sigma^2}{N(N-1)}\sum_{i<j} \mid\Ni_i \cap \Ni_j\mid }_{C_2}+\underbrace{\textstyle  \frac{\mu^2 N}{N-1} (\overline{\mid\Ni\mid^2} - \overline{\mid\Ni\mid}^2) }_{C_3} \bigm\}. \nonumber
\end{flalign}
The right hand side of \eqref{eq:fullrand1} (top) is the mean square error of the difference-in-means estimator due to a complete randomization strategy in the absence of a network, since $(\gamma^2 + \sigma^2)$ is the total variance implied by the network-sum model. 
The three factors $C_{1:3}$ in \eqref{eq:fullrand1} (bottom) can be seen as the contribution to the variance due to the presence of network-correlated outcomes. 
The term $C_1$ is proportional to the average degree of the nodes; thus networks with higher average degrees will tend to lead to higher mean square errors, ceteris paribus.
The term $C_2$ is proportional to the average number of shared neighbors among all pairs of nodes; thus 
networks that are locally denser will tend to have lower mean square error, ceteris paribus.
The term $C_3$ is proportional to the  variance of observed degrees; thus low variability in the degree of the nodes will lead to lower mean square error, ceteris paribus. 
Interestingly, this contribution is not necessarily positive, because of term $C_2$, which summarizes average local density. However, the contribution depends on summaries of the degree distribution of the  network available pre-intervention that are not under control of the designer.


\subsection{More general models of network-correlated potential outcomes}
\label{sec:generalization}

The normal-sum model introduced in Section \ref{sec:normal-sum}, and the normal-mean model introduced in Appendix \ref{sec:nmm-analysis}, are special cases of a more general model which replaces Equation \eqref{eq:nsy0} in the main paper and Equations \eqref{eq:nmy0} in the appendix with the more general formulation,
\begin{equation}
	Y_i(0) \mid X \overset{ind}{\sim} \hbox{ Normal }(g[\{X_j\}_{j\in \Ni_i}], \gamma^2),
\end{equation}
with regularity conditions on the function $g$, to essentially ensuring that, for any subset of nodes  $\mathcal{S} \subset \Ni_i$, the conditional
expectation $E(g[\{X_j\}_{j\in\Ni_i}] | \{X_j\}_{j\in\mathcal{S}})$ is well behaved. We detail the regularity conditions (i.e., positivity, symmetry, and monotonicity) as well as the general form of the mean square error for this more general
model in Appendix \ref{section:general-model}.
In addition, we show that the general form of the mean squared error suggests that good designs 
seek to decrease the number of neighbors shared within treatment groups and increase the number of units
shared between treatment groups, while balancing the size of the groups, as well as the distribution of 
neighborhood sizes. 
These derivations indicate that the network balance criteria the proposed restricted randomizations are based upon extend well beyond the normal-sum model. Moreover, 
model-assisted strategies come with theoretical guarantees that hold regardless of the validity of the
model, as we show next.


\section{Methodology and theory}
\label{sec:tm}



Randomization strategies are probability distribution on the set of binary vectors $\mathcal{Z}$.
Restricted randomization strategies are probability distributions implied by discarding allocation vectors $\W \in \mathcal{Z}$ according to a set of rules. 
We review classical strategies in Section \ref{sec:classicrnd}, introduce new strategies in Section \ref{sec:marr} and study their theoretical properties in Section \ref{sec:theory}. Section~\ref{sec:inference} briefly discusses inference.

\subsection{Classical randomization and restricted randomization strategies}
\label{sec:classicrnd}



According to a Bernoulli randomization 
strategy with parameter $p \in( 0,1)$, each treatment allocation vector $\W \in \mathcal{Z}$ 
 has individual treatments $Z_i$ drawn as independent Bernoulli random variables with probability of success $p$, for $i=1,\dots,N$ units in the population.


A completely randomized design 
with parameters $(N_0,N_1)$, where $N_0+N_1=N$, only considers treatment allocation vectors $\W \in \mathcal{Z}$ such that $\sum_{i-1}^N Z_i =N_1$, and assigns equal probability to them.
If $N_0=N_1=N/2$ we refer to it as a balanced completely randomized design.


Restricted randomization  strategies stem from the observation that, when designing an experiment, it is often clear how to evaluate whether  a treatment allocation vector is undesirable. For instance, when an allocation vector $\W$ leads to statistical imbalance for one or more key covariates, it leaves the door open to confounding even in the presence of randomization \citep{Gosset:1938sf,Cox:1982rr}. 
Indeed, the most common form of restricted randomization is to discard treatment allocations that lead to covariate imbalances
\citep{morgan2012rerandomization}.


\subsection{Model-assisted restricted randomization strategies}
\label{sec:marr}

We introduce four  model-assisted designs, which differ 
 by the degree of reliance on the model; namely, on  the conditional \MSE for the difference-in-means estimator. 

First, we consider balanced restricted randomization strategies, which discard treatment allocation vectors where the number of treated units $N_1$ differs from the number of untreated units $N_0$---or differs by more than one in the case of $N$ odd. This strategy aims at minimizing the contribution of the total variance to the conditional \MSE, according to \eqref{interpretation:var1}.

Second, we introduce unbiased restricted randomization strategies, which discard treatment allocation vectors where the average number of neighbors for treated units differs from the average number of neighbors for untreated units. This strategy aims at minimizing the contribution of the bias to the conditional \MSE, as suggested by the discussion of \eqref{interpretation:bias}.

Third, we introduce optimal restricted randomization strategies, which discard treatment allocation vectors that 
 minimize the average number of shared neighbors among pairs of treated units, according to \eqref{eq:31},
 minimize the average number of shared neighbors among pairs of untreated units, according to \eqref{eq:32},
 and maximize the average number of shared neighbors among pairs of units one of which is treated and the other untreated, according to \eqref{eq:33}.
 
Let $\mathcal{Z}\equiv\{0,1\}^N$ be the set of all possible treatment allocation vector on $N$ units.
Formally, we can define sets of allocations corresponding to the restricted randomization defined above as
\begin{flalign}
 & \textstyle \mathcal{Z}^{b} \equiv \{ \W \in \mathcal{Z}: N_1 - N_0 = 0 \} \label{eq:Zb}\\
 & \textstyle \mathcal{Z}^{u} \equiv \{ \W \in \mathcal{Z}: \frac{1}{N_1} \sum_{\{i:Z_i=1\}} \mid \Ni_i \mid - \frac{1}{N_0} \sum_{\{i:Z_i=0\}} \mid \Ni_i \mid = 0 \} \label{eq:Zu}\\
 & \mathcal{Z}^{o} \equiv \{ \W \in \mathcal{Z}: \mse(\hat{\tau} \mid \W) \leq q_\alpha^{\rm MSE} \}\label{eq:Zo},
\end{flalign}
where $q_\alpha^{\rm MSE}$ is the $\alpha$ quantile of the distribution of the conditional \MSE.
These subsets of randomizations depend on network statistics that the normal-sum model suggests as relevant for computing the conditional \MSE, discussed in Section \ref{sec:cond-mse}.

The rest of the paper focuses on the first three model-assisted strategies:
 balanced restricted randomization design, which assigns equal probability to all $\W \in \mathcal{Z}^b$,
 balanced/unbiased restricted randomization design, which assigns equal probability to all $\W \in \mathcal{Z}^b\cap\mathcal{Z}^u$,
 balanced/unbiased/optimal restricted randomization design, which assigns equal probability to all $\W \in \mathcal{Z}^b\cap\mathcal{Z}^u\cap\mathcal{Z}^o$.
We prove in Appendix~\ref{appendix:proof-thm} that if $\mathcal{Z}^b\cap\mathcal{Z}^u \neq \emptyset$ then $\mathcal{Z}^b\cap\mathcal{Z}^u\cap\mathcal{Z}^o$ contains at least two elements $\W$.



The fourth model-assisted strategy, which we refer to as unconstrained/optimal restricted randomization design, aims at trading off small increases in bias for significant reductions in variance. 
This design assigns equal probability to all $\W\in\mathcal{Z}^{\rm min}$, defined as
\begin{flalign}
 \mathcal{Z}^{\rm min} \equiv \{ \W \in \mathcal{Z}: \arg\min\mse(\hat{\tau} \mid \W)  \}.
\end{flalign}
Even in situations where the set $\mathcal{Z}^{\rm min}$ contains a single allocation vector, because we can only approximately search the space $\mathcal{Z}$ for the optimal vector $\W$ and because we use multiple initializations to perform such a search, in practice  $\mathcal{Z}^{\rm min}$  contains multiple allocation vectors.


\subsection{Model-based optimal treatment allocation strategies}
\label{sec:mbrr}

The four model-assisted strategies in Section \ref{sec:marr} leverage a model for the outcomes for selecting allocations that improve properties of the difference-in-means estimator, which ignores the model. One may wonder why not leveraging the model for the outcomes to also derive a better estimator for the average treatment effect, and then selecting allocations that  improve properties of that estimator. Here, we develop such a model-based optimal treatment allocation strategy.

The natural next step is to replace the difference-in-means estimator with the maximum likelihood  estimator for $\tau$ under the normal-sum model. The estimator $\hat\tau_{\rm mle}$ and its conditional mean square error are derived in Appendix \ref{sec:mle-analysis}.
The optimal maximum likelihood design is then the model-based restricted randomization strategy that assigns equal probability to all $\W\in\mathcal{Z}^{\rm mle}$, defined as
\begin{flalign}
 \mathcal{Z}^{\rm mle} \equiv \{ \W \in \mathcal{Z}: \arg\min\mse(\hat{\tau}_{\rm mle} \mid \W)  \}.
\end{flalign}

In the appendix, we evaluate the performance of the maximum likelihood estimator for $\tau$ using a completely randomized design, as a baseline, to quantify the improvement due to optimal 
restricted randomization.
When evaluating the performance of both model-based strategies, we fix parameters $\mu, \sigma$, and $\gamma$ at their true value, and consider $\tau$ as the only unknown parameter. 


\subsection{Restricted randomizations via rerandomization}
\label{sec:rerandomization}

A general approach for sampling from arbitrary restricted randomization designs, referred to 
as rerandomization, has been recently formalized by \cite{morgan2012rerandomization}. If $\phi$ is a binary function
such that assignment $\W$ is belongs to the restricted randomization set if and only if $\phi(\W) = 1$, then
one simple way to  sample from the restricted randomization design by using a simple rejection sampling approach:
draw an assignment $\W$ from the original design, then keep the assignment if $\phi(\W) = 1$, and
reject it if $\phi(\W) = 0$. In our setting, the restricted sets defined in \ref{eq:Zb}--\ref{eq:Zo}
can be defined terms of different functions $\phi$. Denote the indicator function by $I(.)$, then
\begin{align} 
 & \textstyle \phi^b(\W) = I\bigm\{\sum_i^N Z_i = \sum_i^N (1-Z_i)\bigm\} \\
 & \textstyle \phi^u(\W) = I\bigm\{\mu \cdot \delta_{\Ni}(Z) = 0\bigm\} \\
 & \textstyle \phi^o(\W) = I\bigm\{\mse(\hat{\tau} | \W) \leq q^{MSE}_\alpha\bigm\}.
\end{align} 
Thus rerandomization can be used to sample from the restricted randomization
designs we proposed. Rerandomization as a means to implement restricted randomization strategies is particularly useful when performing exact tests and computing confidence intervals, as detailed next.


\subsection{Inference via inversion of a sequence of exact Fisher tests}
\label{sec:inference}

There are traditionally three  types of confidence intervals in randomization-based inference: 
Neymanian intervals, bootstrap intervals, and Fisher intervals. Neymanian inference in the context of
 restricted randomization is generally a challenging problem. Recent work 
 has proposed an asymptotic theory of re-randomization \citep{Li:2016os}. Unfortunately, the asymptotic 
regime considered there is not compatible with our setting for  two reasons:
(i) proposed methods require the number of covariates to be fixed in the asymptotic regime, while in our case the quantities that are analogous to covariates include the number of neighbors shared by each pair of units, which grows with the number of units in the asymptotic regimes of interest; and (ii) proposed methods also require the constraints to be a function only of the vector of differences in means (between treated and control units) for the observed covariates, and of the variance-covariance matrix of that vector; a condition that does not hold in our case (see appendix).
%
Bootstrap intervals are difficult to implement since the correlation structure of the outcomes may be
complex. 

Instead, we propose using Fisher intervals, which are obtained by inverting a sequence of Fisher exact tests 
\citep[e.g.,][]{rosenbaum2002covariance}. This can be accomplished by means of rerandomization
\citep[e.g., see][sec 2.2]{morgan2012rerandomization}, but by using the proposed restricted randomization distributions as the permutation distributions. Details  are given in Appendix \ref{section:fisher-test}.

To illustrate the potential gains from Fished interval estimates based on restricted randomization, we ran a
 simulation in which we generated outcomes from the normal-sum model, and computed Fisher intervals 
based on balanced optimal restricted randomization (with $\alpha = 0.05$). For a fixed network of 500 nodes,
 we generated two hundred realizations of the potential outcomes 
 according to the normal-sum model, and two hundred observed assignments. For each realization, we computed a Fisher confidence interval based
 on balanced optimal restricted randomization and Fisher intervals based on balanced complete randomization. 
 The results are displayed in Figure~\ref{fig:fisher}. We see that the proposed design based on restricted randomizations reduces the size of the Fisher confidence intervals (at the same level of confidence) compared to the length of the intervals obtained under complete randomization.
The exact design of this simulation, as well as a more extensive simulation and corresponding results, are given in Appendix \ref{section:fisher-test}.

\begin{figure}[t!]
	\centering
	\includegraphics[scale=0.5]{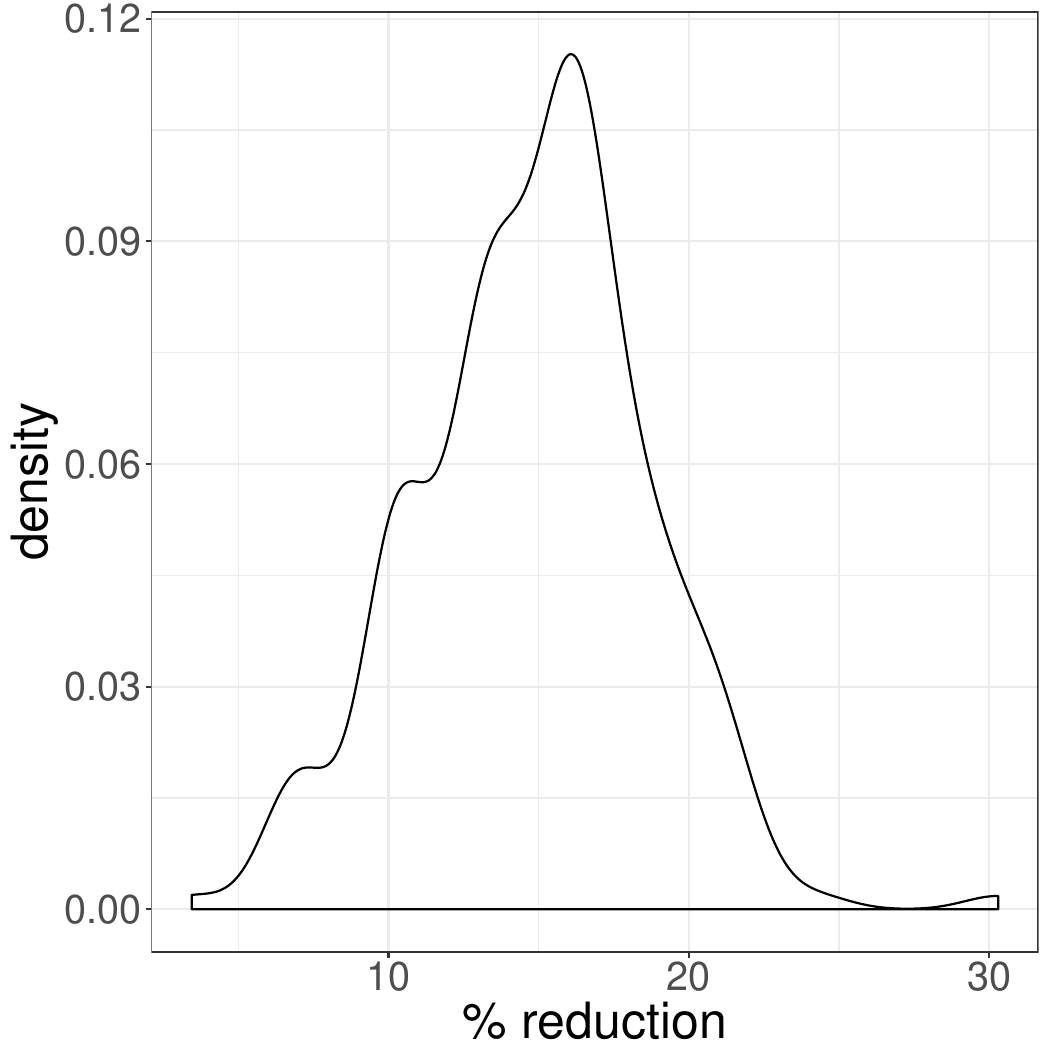}
	\caption{Percent reduction in the length of confidence intervals ($\alpha = 0.05$) obtained with a randomization restricted to $\mathcal{Z}_b \cap \mathcal{Z}_o$, relative to the length of intervals under complete randomization.}
	\label{fig:fisher}
\end{figure}


\subsection{Theory}
\label{sec:theory}

An advantage of model-assisted designs is that
they only partly depend on a model for the outcomes. Thus it is reasonable to expect that these designs might lead to desirable inferential properties even when the model they rely on for evaluating treatment allocations is wrong. 
Following this intuition, here we show that the difference-in-means estimator is design unbiased \citep[e.g., see][]{sarndal2003model} for the restricted randomization strategies developed in Section \ref{sec:marr}.

\begin{definition}[Design unbiasedness]
An estimator $\hat{\tau}$ is unbiased with respect to a distribution on $\mathcal{Z}$, typically referred to as a design on $\mathcal{Z}$, if we have: ~$\sexp_\mathcal{Z}(\hat\tau-\tau)=0$
\end{definition}
The main result follows.
Proofs are given in Appendix \ref{app:proofs}.
\begin{theorem}\label{th:theory}
 The difference-in-means estimator $\hat{\tau}$ defined in \eqref{eq:naive-est} is an unbiased estimator of the average treatment effect  with respect to the following distributions:
\vspace{-6pt}
 \begin{itemize}
  \item[(i)] Uniform distribution on $\mathcal{Z}^b$, which defines the balanced  design
  \item[(ii)] Uniform distribution on $\mathcal{Z}^b\cap\mathcal{Z}^u$, which defines the balanced/unbiased  design
  \item[(iii)] Uniform distribution on $\mathcal{Z}^b\cap\mathcal{Z}^o$, which defines the balanced/optimal design
  \item[(iv)] Uniform distribution on $\mathcal{Z}^b\cap\mathcal{Z}^u\cap\mathcal{Z}^o$, which defines the balanced/unbiased/optimal.    
 \end{itemize}	
\end{theorem}
Intuitively, as a consequence of design unbiasedness and of the increasingly nested supports, we can compare variances of $\hat\tau$ implied by the designs in Theorem \ref{th:theory}, in expectation. 
\begin{corollary}\label{cor:theory}
Let $\hat{\tau}$ be the  estimator defined in \eqref{eq:naive-est}. We have,
\begin{equation*}
	\sexp\bigm\{\svar_{\mathcal{Z}^b\cap\mathcal{Z}^o}( \hat{\tau} \mid \Y)\bigm\} \,\,\leq~ \sexp\bigm\{\svar_{\mathcal{Z}^b}( \hat{\tau} \mid \Y)\bigm\}.
\end{equation*}
\end{corollary}
And similar inequalities can be derived easily for any pairs of nested designs in Theorem \ref{th:theory}.

These results are based on symmetry arguments, which is why $\mathcal{Z}^b$ is always part of the support of designs that make the difference-in-means estimator unbiased. This notion of symmetry  is made precise in the following Lemma.
\begin{lemma}\label{lemma:trick}
For $\W$ in $\mathcal{Z}^b$, we have: ~$\hat{\tau}(1-\W) = 2 \tau - \hat{\tau}(\W)$.
\end{lemma}
As a consequence, if we required the  unconstrained optimal design to be balanced, by restricting its support to $\mathcal{Z}^b\cap\mathcal{Z}^{\rm min}$, we would recover design unbiasedness for the difference-in-means estimator. However, we do not consider balanced unconstrained optimal designs.
\vspace{-3pt}



\section{Discussion}


%
%
In this paper we have introduced a strategy for model-assisted design of experiments.
Given the difference-in-means estimator, we use a model for the outcomes to compute its mean squared error conditional on a fixed treatment allocation vector $\W\in\mathcal{Z}$. This calculation  identifies  network statistics that are relevant in controlling bias and variance terms. We then restrict the support of the randomization distribution to specific subsets of $\mathcal{Z}$ that minimize some of these terms. Should the model not hold, the difference-in-means estimator remains unbiased for the average treatment effects with respect to the restricted randomization distribution, as detailed in Theorem \ref{th:theory}.
In the model-assisted survey sampling literature, in contrast, given a linear weighted estimator such as the Horwitz--Thompson, a model is used to derive correction factors for the weights. The corrected estimator has reduced variance if the model holds, and is otherwise unbiased with respect to the sampling distribution independently of the model. 

%

The idea behind model-assisted design is fairly general, two key elements being the estimator and the model.
The theoretical guarantees  in Section~\ref{sec:theory} are limited to estimators satisfying the symmetry condition of Lemma~\ref{lemma:trick}, and to the model family in Appendix \ref{section:general-model}.
%
Extending the theory to a larger class of estimators and models is conceptually feasible, although often results in complex expressions for the mean square error, and  hard-to-interpret balance criteria.

In practice, there are often additional issues to consider, which we have assumed away in our analysis for simplicity of exposition. Importantly, 
 covariates will have to be taken into account, 
 and the parameters $\mu, \sigma^2$, and $\gamma$ will need to be specified or estimated.
Options for inference on the parameters include point priors \citep{box1959design}, or specifying full priors to work with the integrated mean squared error. In both situations, historical data and pilot studies might be used to calibrate these priors, and are recommended for optimal design in practice \citep{kim2015aa,shakya2016aa}.
The theory we developed for a more general model of network-correlated outcomes, and extensive simulation studies, both detailed in the Appendix, show that the gains in terms of efficiency one can expect to achieve with model-assisted design of experiments (over design-based and model-based strategies) are  robust to a large degree of misspecifications.

%
%

This paper is a starting point.
In the context of the literature on homophily and peer-influence, this paper suggests a viable strategy to get an analytical handle on which features of a network might be useful to control when designing an experiment. 
However, we limit ourselves to the case of network-correlated outcomes in the absence of peer-influence, we only analyze the conditional mean square error for the differences-in-means estimator under the normal-sum model, in Sections \ref{sec:nsm-analysis} and \ref{sec:tm}, and under the normal-mean model, in Appendix\ref{sec:nmm-analysis}, the conditional mean square error for the maximum likelihood estimator, in Appendix \ref{sec:mle-analysis}, and we carry out an empirical sensitivity analysis.
We initially choose to tackle network-correlated outcomes in isolation to gain clear analytical insights. We are currently working on combining these insights to design randomization strategies that can optimally estimate causal effects of interest in the presence of both network interference and confounding due to network correlations. 


%

\bibliographystyle{plainnat}
\bibliography{doe}

\newpage
\fontsize{10pt}{12pt}\selectfont

\section*{Supplementary Material}

The supplement is organized as follows. 
Section~\ref{section:general-model} provides an in-depth analysis of the general model mentioned in the article, including all proofs. 
Section~\ref{section:fisher-test} provides a detailed procedure
for inverting Fisher exact tests and obtaining confidence intervals with restricted randomization
designs. It also reports results of an extended simulation study comparing the size of the confidence intervals obtained by inverting Fisher exact tests based on restricted randomizations to the size of confidence intervals obtained by inverting Fisher exact tests based on complete randomization. 
Section~\ref{section:results} presents the results of extensive simulation studies illustrating the robustness of our model-assisted strategies to model-misspecification, in contrast with the model-based strategies.
Section~\ref{sec:derivations} presents detailed derivations and proofs of theorems and lemmas.

\appendix




\section{More general models of network-correlated outcomes}
\label{section:general-model}

In this section, we introduce a more general class of models, for which the interpretation of the conditional mean square error---used for the purpose of restricting randomizations in the model-assisted design strategies---generalizes that of the normal-sum and normal-mean models. This illustrates the extent to which the design guidelines derived from the simpler models hold more generally.
 Section~\ref{subsection:new-results} introduces the general model formulation and states the results for it. Section~\ref{subsection:examples} gives examples of different instances of this model. Section~\ref{subsection:new-results-proofs} details the proofs.

\subsection{Elements of experimental design for more general models}
\label{subsection:new-results}

Although the exact network balance criteria will depend on the model, some broad experimental design guidelines are available for a relatively large class of models. Consider the family,
\begin{flalign*}
	Y_i(0) \vert \textbf{X} &\overset{ind}{\sim} \mathcal{N}\bigg( g( \{X_j\}_{j\in \mathcal{N}_i}), \gamma^2 \bigg) \\
	X_i &\overset{iid}{\sim} \mathcal{N}(\mu,  \sigma^2) \\
	Y_i(1) &= Y_i(0) + \tau
\end{flalign*}
where the $g$ satisfies the following regularity conditions:
\begin{constraint}\label{constr}
	Let $\{X_1, \ldots, \}$ a sequence of iid random variables. There exists a real-valued set function $\phi$ 
	and a real-valued function of three variables $h(\cdot, \cdot, \cdot)$ such that for any collection $\{X_k\}_{k\in \chi}$
	indexed by a finite set $\chi$ and any subset of indices $\Scal \subset \chi$, the following hold:
	\begin{enumerate}
		\item $\sexp[ g(\{X_k\}_{k\in \chi}) | \{X_k\}_{k\in \Scal} ] = h(|\chi|, |\mathcal{S}|, \phi(\{X_k\}_{k\in \Scal}))$
		\item $h(n, s, \cdot)$ is a monotone function of its third argument, for $n$ and $s$ fixed.
		That is, $h(n, s, \cdot)$ is either non-increasing for every $n$ and $s$, or non-decreasing for 
		every $n$ and $s$.
		\item If $s_0 \in \chi \cap \bar{\Scal}$, the quantity $\phi(\{X_{s_0} \cup \{X_k\}_{k\in \Scal}\})$ is a non-decreasing
		function of $X_{s_0}$, for $\{X_k\}_{k\in \Scal}$ fixed.
	\end{enumerate}
\end{constraint}

We now state our main new theorems. We will give examples of functions $g$ satisfying the constraints immediately after:

\begin{theorem}
If the regularity conditions above hold, we have:
\begin{equation*}
\mbox{MSE}(\hat{\tau} | \W) = ( \delta(\W) )^2 + \gamma^2 \omega(\W)' \cdot \omega(\W) + \omega(\W)' \cdot \Sigma \cdot \omega(\W)
\end{equation*}
where:
\begin{equation*}
	\delta(\W) = \frac{1}{N_1} \sum_{i:Z_1} q(|\Ni_i|) - \frac{1}{N_0} \sum_{i:Z_i=0} q(|\Ni_i|)
\end{equation*}
and:
\begin{equation*}
\gamma^2 \omega(\W)' \cdot \omega(\W) = \gamma^2 \bigg( \frac{1}{N_1} + \frac{1}{N_0} \bigg)
\end{equation*}
and:
\begin{flalign*}
\omega(\W)' \cdot \Sigma \cdot \omega(\W) &= \frac{1}{N_1^2} \sum_{i:j:Z_i=Z_j=1} m(|\Ni_i|, |\Ni_j|, |\Ni_i\cap\Ni_j|) \\
&+ \frac{1}{N_0^2} \sum_{i:j:Z_i=Z_j=0} m(|\Ni_i|, |\Ni_j|, |\Ni_i\cap\Ni_j|) \\
&- \frac{2}{N_1 \cdot N_0}  \sum_{i:j:Z_i= 1 \, \mbox{ and } \, Z_j=0} m(|\Ni_i|, |\Ni_j|, |\Ni_i\cap\Ni_j|) \\
&+ \frac{1}{N_1^2} \sum_{i:Z_i=1} v(|\Ni_i|) + \frac{1}{N_0^2} \sum_{i:Z_i=0}  v(|\Ni_i|) 
\end{flalign*}
where $q$, $v$ and $m$ all depend on $g$, $\mu$, and $\sigma^2$ and are defined in Lemma \ref{lemma:basics}, below. In addition,
the function $m$ has the following properties:
\begin{enumerate}
	\item $m(a,b,c) \geq 0$ for all $a, b \geq 0$ and $ c \leq min(a, b)$.
	\item $m$ is symmetric in its first two arguments: $m(a, b, c) = m(b, a, c)$
	\item $m$ is a non-decreasing function of its third argument, when the first two
	arguments are held constant.
\end{enumerate}
\end{theorem}

\begin{remark}
	The functions $q$ and $h$ in the theorem depend on the specific choice of $g$, and on the 
	parameters. However when looking at the variance one can see that everything else 
	being equal, the mean square error is minimized by decreasing within-group overlap, and 
	by increasing between-group overlap, precisely as in the basic normal-sum model. A general 
	heuristic for experimental design would then be: minimize shared neighbors within treatment and  
	control groups, maximize shared neighbors between these groups, while keeping the distribution of 
	neighborhood sizes similar in both treatment and control groups.
\end{remark}
%

\subsection{Example models}
\label{subsection:examples}

We now give examples of functions $g$ satisfying the three regularity conditions stated above.
\begin{example}
	Consider $g(\{X_j\}_{\Ni_i}) = \sum_{j \in \Ni_i} X_j$. Let $\Scal \subset \Ni_i$. We
	have:
	\begin{equation*}
		\sexp[ g(\{X_j\}_{\Ni_i}) | \{X_k\}_{k\in \Scal} ] = (|\Ni_i| - |\Scal|) \mu + \sum_{k\in \Scal} X_k = h(|\Ni_i|, |\Scal |, \phi(\{X_k\}_{k\in\Scal}))
	\end{equation*}
	with $\phi( \{X_k\}_{k\in \Scal} ) = \sum_{k\in \Scal} X_k$. It is clear that $h(n, s, x)$ is an increasing function of $x$ for fixed $n$ and $s$,
	and that $\phi(\{X_k\}_{k\in \Scal})$ is a non-decreasing function of any element of the set, after fixing the others.
	So $g(\{X_j\}_{\Ni_i}) = \sum_{j \in \Ni_i} X_i$ satisfies all three regularity conditions.
\end{example}

\begin{example}
	Consider $g(\{X_j\}_{\Ni_i}) = \iv( \sum_{j \in \Ni_i} X_j > c)$. Let $\Scal = \subset \Ni_i$.
	With regards to the first regularity condition we have:
	\begin{flalign*}
		\sexp[ g(\{X_j\}_{\Ni_i}) | \{X_k\}_{k\in\Scal} ] &= P( \sum_{j \in \Ni - \Scal} X_j > c - \sum_{k\in \Scal} X_k ) \\
		&= 1 - \Phi\bigg( \frac{  (c- \sum_{j\in \Scal} X_j) - (|\Ni_i| - |\Scal |)\mu }{(|\Ni_i| - |\Scal|) \sigma^2} \bigg) \\
		&= h(|\Ni_i|, \phi(\Scal))
	\end{flalign*}
	where $\phi( \{X_k\}_{k\in \Scal} ) = \sum_{S\in \Scal} S$. The other two regularity conditions are easily verifiable.\bigskip
\end{example}

\begin{example}
	Consider $g(\{X_j\}_{\Ni_i}) = \mbox{max}(\{X_j\}_{\Ni_i})$. As before, let $\Scal \subset \Ni_i$. 
	Also, let $\bar{\Scal} = \Ni_i -  \Scal$.
	\begin{flalign*}
		\sexp[g(\{X_j\}_{\Ni_i}) \vert \{X_k\}_{k\in \Scal}] 
		&= \sexp\bigg[ \iv\bigg(\max( \{X_k\}_{k \in \bar{\Scal}}) > \max( \{X_k\}_{k \in \Scal} )\bigg) \max(\{X_k\}_{\bar{\Scal}}) \\
		&+ \iv\bigg(\max( \{X_k\}_{k\in \bar{\Scal}}) < \max( \{X_k\}_{k\in \Scal} )\bigg) \max\bigg(\{X_k\}_{k\in\Scal}\bigg) \bigg\vert \{X_k\}_\Scal \bigg] \\
		&= P\bigg(\max( \{X_k\}_{\bar{\Scal}}) > \max( \{X_k\}_{\Scal} ) \bigg\vert \{X_k\}_{\Scal} \bigg) \sexp \bigg[\max(\{X_k\}_{\bar{\Scal}}) \vert   \max(\{X_k\}_{\bar{\Scal}}) > \max( \{X_k\}_{\Scal} )\bigg] \\
		&+ P\bigg(\max( \{X_k\}_{\bar{\Scal}}) < \max( \{X_k\}_\Scal ) \vert \{X_k\}_\Scal \bigg) \max(\{X_k\}_\Scal) \\
		= h(|\Ni_i|, |\Scal|, \phi(\{X_k\}_\Scal))
	\end{flalign*}
	where $\phi(\{X_k\}_\Scal) = \max(\{X_k\}_\Scal)$. We have used the fact that the distribution of $\max(\bar{\Scal})$ depends on $\Scal$ and $\Ni_i$ only through $|\Ni_i| - |\Scal|$, since the
	$X_j$'s are iid. This means that $P\bigg(\max( \{X_k\}_{\bar{\Scal}}) > \max( \{X_k\}_\Scal ) \vert \{X_k\}_\Scal \bigg)$ is a function of $|\Ni_i|$ and $|\Scal|$ and $\max(\{X_k\}_\Scal)$.
	It is easy to verify the last two regularity conditions are also satisfied.\bigskip
\end{example}

\subsection{Proofs}
\label{subsection:new-results-proofs}

\begin{lemma}\label{lemma:cov}
	If the regularity conditions hold, then for any $X_1, \ldots $ iid random variables
	 there exists a function $m(\cdot, \cdot, \cdot)$ such that for any $\chi_1$ and $\chi_2$ two sets of indices 
	such that $\chi_1 \neq \chi_2$, and such that $\Scal = \chi_1 \cap \chi_2 \neq \emptyset$, 
	we have the following:
	\begin{equation*}
	\scov\bigg[g(\{X_k\}_{k\in \chi_1}), g(\{X_k\}_{k \in \chi_2})\bigg] = m(|\chi_1|, |\chi_2|, |\Scal |)
	\end{equation*}
	and the function $m$ satisfies:
	\begin{enumerate}
		\item (positivity) $m(a,b,c) \geq 0$ for all $a, b \geq 0$ and $ c \leq min(a, b)$
		\item (partial symmetry) $m(a, b, c) = m(b, a, c)$
		\item (monotonicity) $m$ is a non-decreasing function of its third argument. That is, for all $a$ and $b$,
		\begin{equation*}
		m: c \rightarrow m(a, b, c)
		\end{equation*}
		is a non-decreasing function.
	\end{enumerate}
\end{lemma}

\begin{proof}
Let $X_1, \ldots $ iid random variables, $\chi_1$ and $\chi_2$ two sets of indices such that
$\chi_1 \neq \chi_2$ and $S=\chi_1 \cap \chi_2 \neq \emptyset$. We have:
\begin{flalign*}
	\scov\bigg[g(\{X_k\}_{k\in \chi_1}), g(\{X_k\}_{k\in \chi_2})\bigg] &= \sexp\bigg[ \scov\bigg(g(\{X_k\}_{k\in \chi_1}), g(\{X_k\}_{k\in \chi_2})\bigg) \vert \{X_k\}_{k\in \Scal}\bigg] \\
	&+ \scov\bigg[\sexp[ g(\{X_k\}_{k\in \chi_1})| \{X_k\}_{k\in \Scal}], \sexp[g(\{X_k\}_{k\in \chi_2}) | \{X_k\}_{k\in \Scal}] \bigg] \\
	&= 0 + \scov\bigg[\sexp[ g(\{X_k\}_{k\in \chi_1}) | \{X_k\}_{k\in \Scal}], \sexp[g(\{X_k\}_{k\in \chi_2}) | \{X_k\}_{k\in \Scal}] \bigg] \\
	&= \scov\bigg[ h(|\chi_1|, \phi( \{X_k\}_{k\in \Scal})), h(|\chi_2|,  \phi(\{X_k\}_{k\in \Scal})) \bigg]
\end{flalign*}
where the last equality uses the first regularity condition. Since the X's are iid, the covariance depends on $\{X_k\}_{k\in \Scal}$
only through $|\Scal|$, the number of random variables in the set, and so we can write:
\begin{flalign*}
\scov\bigg[g(\{X_k\}_{k\in \chi_1}), g(\{X_k\}_{k\in \chi_2})\bigg] &= \scov\bigg[ h(|\chi_1|, |\Scal|, \phi( \{X_k\}_{k\in \Scal})), h(|\chi_2|, |\Scal|, \phi(\{X_k\}_{k\in \Scal})) \bigg] \\
&= m(|\chi_1|, |\chi_2|, |\Scal|)
\end{flalign*}

where we emphasize once again that implicitly, $m$ depends on $g$ and on the model. We now study the properties of $m$:

\medskip

\noindent\textit{Partial Symmetry:}
Since the covariance is symmetric, it is clear that:
\begin{equation*}
m(|\chi_1|, |\chi_2|, |\Scal|) = m(|\chi_2|, |\chi_1|, |\Scal|)
\end{equation*}

\medskip

\noindent\textit{Positivity:}
By the second regularity condition, $h(n, \cdot)$ is either a non-decreasing function of its second argument for all $n$, or a non-increasing
function of its second argument for all $n$. But it is known (see e.g \cite{thorisson1995coupling}, section 2) that the covariance of two monotone 
functions of random variables is positive. Thus:
\begin{equation*}
	\scov\bigg[ h(|\chi_1|, \phi( \{X_k\}_{k\in \Scal})), h(|\chi_2|,  \phi(\{X_k\}_{k\in \Scal})) \bigg] \geq 0
\end{equation*}
and we have:
\begin{equation*}
	m(|\chi_1|, |\chi_2|, |\Scal|) \geq 0.
\end{equation*}

\medskip

\noindent\textit{Monotonicity:}
Let $\chi_1'$ and $\chi_2'$ such that $|\chi'_1| = |\chi_1|$ and $|\chi'_2| = |\chi_2|$ and 
$\chi'_1 \cap \chi'_2 = \{s_0\} \cup \Scal \equiv \Scal'$. On the one hand, we have:
\begin{equation}\label{eq:one-hand}
	\scov\bigg[g(\{X_k\}_{k\in \chi'_1}), g(\{X_k\}_{k\in \chi'_2})\bigg]  =  
	\scov\bigg[ h(|\chi'_1|, \phi( \{X_k\}_{k\in \Scal'})), h(|\chi'_2|,  \phi(\{X_k\}_{k\in \Scal'})) \bigg]
\end{equation}
on the other hand, we also have:
\begin{flalign*}
	\scov\bigg[g(\{X_k\}_{k\in \chi'_1}), g(\{X_k\}_{k\in \chi'_2})\bigg] &= \sexp\bigg[ \scov[g(\{X_k\}_{k\in \chi'_1}), g(\{X_k\}_{k\in \chi'_2})\vert \{X_k\}_{k\in \Scal} ]\bigg] \\
	&+ \scov\bigg[\sexp[ g(\{X_k\}_{k\in \chi'_1})| \{X_k\}_{k\in \Scal}], \sexp[g(\{X_k\}_{k\in \chi'_2}) | \{X_k\}_{k\in \Scal}] \bigg] \\
	&= \sexp\bigg[  \scov[g(\{X_k\}_{k\in \chi'_1}), g(\{X_k\}_{k\in \chi'_2})\vert \{X_k\}_{k\in \Scal}] \bigg] \\
	&+ \scov[  h(|\chi'_1|, \phi(\{X_k\}_{k\in \Scal}) ), h(|\chi'_2|, \phi(\{X_k\}_{k\in \Scal}) ) ] \\
\end{flalign*}
But we have:
\begin{flalign*}
\scov\bigg[g(\{X_k\}_{k\in \chi'_1}), g(\{X_k\}_{k\in \chi'_2})\vert \{X_k\}_{k\in \Scal}\bigg] 
&= \sexp\bigg[ Cov\bigg(g(\{X_k\}_{k\in \chi'_1}), g(\{X_k\}_{k\in \chi'_2})\vert \{X_k\}_{k\in \Scal'}\bigg) \bigg\vert \{X_k\}_{k\in \Scal}\bigg] \\
&+ \scov\bigg[  \sexp\bigg[ g(\{X_k\}_{k\in \chi'_1}) \vert \{X_k\}_{k\in \Scal'} \bigg], \sexp\bigg[ g(\{X_k\}_{k\in \chi'_2}) \vert \{X_k\}_{k\in \Scal'} \bigg] 
\bigg\vert  \{X_k\}_{k\in \Scal} \bigg] \\
&= \scov[  h(|\chi'_1|, \phi(\{X_k\}_{k\in \Scal'}) ), h(|\chi'_2|, \phi(\{X_k\}_{k\in \Scal'}) | \{X_k\}_{k\in \Scal}) ]
\end{flalign*}
The only random element in the covariance of the last line is $X_{s_0}$, since we condition on all the other
random variables. But by the third regularity condition, the function:
\begin{equation*}
	X_{s_0} \rightarrow \phi(\{X_{s_0} \cup \{X_k\}_{k\in \Scal} )
\end{equation*}
for fixed $\{X_k\}_{k\in \Scal}$ is non-decreasing. Thus as above the covariance will be positive (\cite{thorisson1995coupling}), and so:
\begin{equation*}
	\scov\bigg[g(\{X_k\}_{k\in \chi'_1}), g(\{X_k\}_{k\in \chi'_2})\vert \{X_k\}_{k\in \Scal}\bigg]  \geq 0
\end{equation*}
and so we also have:
\begin{equation*}
\sexp\bigg[  \scov[g(\{X_k\}_{k\in \chi'_1}), g(\{X_k\}_{k\in \chi'_2})\vert \{X_k\}_{k\in \Scal}] \bigg] \geq 0
\end{equation*}
and putting it all together, we have:
\begin{equation*}
\scov\bigg[g(\{X_k\}_{k\in \chi'_1}), g(\{X_k\}_{k\in \chi'_2})\bigg] \geq  \scov\bigg[  h(|\chi'_1|, \phi(\{X_k\}_{k\in \Scal}) ), h(|\chi'_2|, \phi(\{X_k\}_{k\in \Scal}) ) \bigg]
\end{equation*}
and combining with Equation~\ref{eq:one-hand}, we have:
\begin{equation*}
	\scov\bigg[ h(|\chi'_1|, \phi( \{X_k\}_{k\in \Scal'})), h(|\chi'_2|,  \phi(\{X_k\}_{k\in \Scal'})) \bigg] \geq
	\scov\bigg[  h(|\chi'_1|, \phi(\{X_k\}_{k\in \Scal}) ), h(|\chi'_2|, \phi(\{X_k\}_{k\in \Scal}) ) \bigg]
\end{equation*}
that is, 
\begin{equation*}
	m(|\chi_1|, |\chi_2|, |\Scal| + 1) \geq m(|\chi_1|, |\chi_2|, |\Scal|)
\end{equation*}
and so by induction, $m$ is a non-decreasing function of its third argument.
\end{proof}

\begin{lemma}\label{lemma:basics}
	If $g$ satisfies the regularity conditions, we have:
	\begin{flalign*}
		\sexpy[Y_i(Z_i) \vert \W] &= \tau Z_i + q( |\Ni_i| ) \\
		\svary[Y_i(Z_i) \vert \W] &= \gamma^2 + v(|\Ni_i|)\\
		\scovy[Y_i(Z_i), Y_j(Z_j) \vert \W] &= m(|\Ni_i|, |\Ni_j|, |\Ni_i \cap \Ni_j|) \quad i \neq j
	\end{flalign*}
	where:
	\begin{flalign*}
		q(|\Ni_i|) &= \sexpy\bigg[ g(\{X_k\}_{k\in \Ni_i}) \bigg\vert \W \bigg] \\
		v(|\Ni_i|) &= \svary\bigg[ g(\{X_k\}_{k\in \Ni_i}) \bigg\vert \W \bigg] \\
	\end{flalign*}
	and:
	\begin{equation*}
	m(|\Ni_i|, |\Ni_j|, |\Ni_i \cap \Ni_j|) = \scovy\bigg[ h\bigg(|\Ni_i|, \phi(\{X_k\}_{k\in \Ni_i \cap \Ni_j})\bigg), 
		h\bigg(|\Ni_j|, \phi(\{X_k\}_{k\in \Ni_i \cap \Ni_j})\bigg) \bigg\vert \W \bigg]
	\end{equation*}
	if $|\Ni_i \cap \Ni_j| \neq 0$ and $m(|\Ni_i|, |\Ni_j|, 0) \equiv 0$, where $q$, $v$ and $m$ depend
	implicitly on $g$ and on the model for $Xs$.
	Moreover, $m$ satisfies the following properties:
	\begin{enumerate}
		\item (positivity) $m(a,b,c) \geq 0$ for all $a, b \geq 0$ and $ c \leq min(a, b)$.
		\item (partial symmetry) $m$ is symmetric in its first two arguments: $m(a, b, c) = m(b, a, c)$
		\item (monotonicity) $m$ is a non-decreasing function of its third argument, when the first two
		arguments are held constant.
	\end{enumerate}
\end{lemma}

\begin{proof}
We proceed in order:

\medskip

\noindent \textit{Expectation:}
\begin{flalign*}
	\sexpy[ Y_i(Z_i) \vert \W] &= \sexpy[ Y_i(0) + Z_i\tau \vert | \W ] \\
	&= \sexpy[ \sexpy[ Y_i(0) | \textbf{X}, \W] \vert \W] + \tau Z_i \\
	&= \sexpy[ g(\{X_k\}_{k\in \Ni_i}) \vert \W] + \tau Z_i
\end{flalign*}

\medskip

\noindent \textit{Variance:}
\begin{flalign*}
	\svary[Y_i(Z_i) \vert \W[ &= \svary[ Z_i \tau + Y_i(0) \vert \W] \\
	&= \sexpy[ Var( Y_i(0) | \textbf{X}, \W) \vert \W ] + \svary[ \sexpy[ Y_i(0) | \textbf{X}, \W] \vert \W] \\
	&= \sexpy[ \gamma^2 \vert \W ] + \svary[ g(\{X_k\}_{k\in \Ni_i}) \vert \W] \\
	&= \gamma^2 + \svary[ g(\{X_k\}_{k\in \Ni_i}) \vert \W]
\end{flalign*}

\medskip 

\noindent \textit{Covariance:} 
\begin{flalign*}
	\scovy[Y_i(Z_i), Y_j(Z_j) \vert \W] &= \scovy[Z_i \tau + Y_i(0), Z_j \tau + Y_j(0) \vert \W] \\
	&= \sexpy\bigg[ \scovy[ Y_i(0), Y_j(0) |\textbf{X} , \W] \bigg \vert \W \bigg] \\
	&+ \scovy\bigg[\sexpy[ Y_i(0) \vert \textbf{X}, \W] ,  \sexpy[ Y_j(0) \vert \textbf{X}, \W] \bigg\vert \W \bigg] \\
	&= \scovy\bigg[g(\{X_k\}_{k\in \Ni_i}),  g(\{X_k\}_{k\in \Ni_j})  \bigg\vert \W \bigg] \\
\end{flalign*}
We now apply Lemma~\ref{lemma:cov} to the RHS of the last equality, with $\chi_1 = \Ni_i$, $\chi_2 = \Ni_j$,
and $\Scal =  \Ni_i \cap \Ni_j $, and we immediately obtain:
\begin{equation*}
\scovy\bigg[g(\{X_k\}_{k\in \Ni_i}),  g(\{X_k\}_{k\in \Ni_j})  \bigg\vert \W \bigg] = \begin{cases}
	m(|\Ni_i|, |\Ni_j|, |\Ni_i \cap \Ni_j|) & \quad \mbox{ if } \, \Ni_i \cap \Ni_j \neq \emptyset \\
	0 & \quad \mbox{ otherwise}
\end{cases}
\end{equation*}
and the properties of $m$ are also obtained from Lemma~\ref{lemma:cov}.
\end{proof}



\begin{proof}[Proof of Theorem]

Recall that:
\begin{equation*}
	\hat{\tau} = \frac{1}{N_1} \sum_{i:Z_i=1} Y_i(1) - \frac{1}{N_0} \sum_{i:Z_i=0} Y_i(0)
\end{equation*}
We have:
\begin{flalign*}
	\bias(\hat{\tau}, \tau | \W) &= \sexpy[ \hat{\tau} | \W] - \tau \\
	&= \frac{1}{N_1} \sum_{i:Z_i = 1} \sexpy[Y_i(1) \vert \W] - \frac{1}{N_0} \sum_{i:Z_i=0} \sexpy[Y_i(0) \vert \W] - \tau \\
	&= \frac{1}{N_1} \sum_{i:Z_i = 1} (\tau + q(|\Ni_i|))  - \frac{1}{N_0} \sum_{i:Z_i = 0} q(|\Ni_i|) - \tau \\
	&= \frac{1}{N_1} \sum_{i:Z_i = 1} q(|\Ni_i|) - \frac{1}{N_0} \sum_{i:Z_i = 0} q(|\Ni_i|)
\end{flalign*}
where we have used Lemma~\ref{lemma:basics}. We now turn to the variance. Let 
$\omega(Z_i) = \frac{Z_i}{N_1} - \frac{1-Z_i}{N_0}$. We have:
\begin{flalign*}
	\svary[\hat{\tau} | \W] &= \svary\bigg[ \sum_i \omega(Z_i) \cdot  Y_i(0) + \tau\bigg] \\
	&= \sum_i \omega(Z_i)^2 \cdot \svary(Y_i(0) \vert \W) + \sum_{i\neq j} \omega(Z_i) \cdot \omega(Z_j) \cdot \scovy(Y_i(0), Y_j(0) \vert \W)
\end{flalign*}
But we have:
\begin{equation*}
	\omega(Z_i)^2 = \begin{cases}
	\frac{1}{N_1}^2 &\quad \mbox{if} \, Z_i = 1 \\
	\frac{1}{N_0}^2 &\quad \mbox{if} \, Z_i = 0
	\end{cases}
\end{equation*}
Applying Lemma~\ref{lemma:basics} to the first term of the RHS of the last line gives:
\begin{flalign*}
\sum_i \omega(Z_i)^2 \cdot \svary(Y_i(0) \vert \W) &= \sum_i \omega(Z_i)^2 (\gamma^2 + v(|\Ni_i|) )\\
&= \sum_{i: Z_i=1} \omega(1)^2 \gamma^2 + \sum_{i: Z_i=0} \omega(0)^2 \gamma^2 + \sum_{i: Z_i=1} \omega(1)^2 v(|\Ni_i|)  + \sum_{i: Z_i=0} \omega(0)^2 v(|\Ni_i|) \\
&= \gamma^2 \cdot \bigg( \frac{1}{N_1} + \frac{1}{N_0}\bigg) + \bigg( \frac{1}{N_1^2} \sum_{i: Z_i=1}  v(|\Ni_i|) + \frac{1}{N_0^2} \sum_{i: Z_i=0}  v(|\Ni_i|) \bigg)
\end{flalign*}
Now notice that:
\begin{equation*}
	\omega(Z_i) \cdot \omega(Z_j) = \begin{cases}
		\frac{1}{N_1^2} &\quad \mbox{if} \, Z_i = Z_j = 1 \\
		\frac{1}{N_0^2} &\quad \mbox{if} \, Z_i = Z_j = 0 \\
		- \frac{1}{N_1N_0} &\quad \mbox{otherwise}
	\end{cases}
\end{equation*}
So applying Lemma~\ref{lemma:basics} to the second term of the RHS of the last line gives:
\begin{flalign*}
\sum_{i\neq j} \omega(Z_i) \cdot \omega(Z_j) \cdot \scov(Y_i(0), Y_j(0) \vert \W) 
&= \sum_{i\neq j} \omega(Z_i) \cdot \omega(Z_j) \cdot  m(|\Ni_i|, |\Ni_j|, |\Ni_i \cap \Ni_j|) \\
& = \frac{1}{N_1^2}\sum_{i\neq j: Z_i = Z_j=1}m(|\Ni_i|, |\Ni_j|, |\Ni_i \cap \Ni_j|) \\
&+ \frac{1}{N_0^2}\sum_{i\neq j: Z_i = Z_j=0}m(|\Ni_i|, |\Ni_j|, |\Ni_i \cap \Ni_j|) \\
&- \frac{2}{N_1N_0} \sum_{i\neq j: Z_i = 1 \, \mbox{ and } \, Z_j=0}m(|\Ni_i|, |\Ni_j|, |\Ni_i \cap \Ni_j|) 
\end{flalign*}
where the $\frac{-2}{N_1N_0}$ term is obtained by symmetry of $m$ with respect to its first two argument.\\

\noindent Now using the fact that $\mse_{\Theta}(\hat{\tau}, \tau | \W) = \bias_{\Theta}(\hat{\tau}, \tau | \W)^2 + \svary[\hat{\tau} | \W]$, we have:
\begin{flalign*}
\mse_{\Theta}(\hat{\tau}, \tau | \W) &= \bigg(\frac{1}{N_1} \sum_{i:Z_i = 1} q(|\Ni_i|) - \frac{1}{N_0} \sum_{i:Z_i = 0} q(|\Ni_i|)\bigg)^2 \\
&+ \gamma^2 \cdot \bigg( \frac{1}{N_1} + \frac{1}{N_0}\bigg)  \\
&+ \frac{1}{N_1^2}\sum_{i\neq j: Z_i = Z_j=1}m(|\Ni_i|, |\Ni_j|, |\Ni_i \cap \Ni_j|) \\
&+ \frac{1}{N_0^2}\sum_{i\neq j: Z_i = Z_j=0}m(|\Ni_i|, |\Ni_j|, |\Ni_i \cap \Ni_j|) \\
&- \frac{2}{N_1N_0} \sum_{i\neq j: Z_i = 1 \, \mbox{ and } \, Z_j=0}m(|\Ni_i|, |\Ni_j|, |\Ni_i \cap \Ni_j|) \\
&+ \bigg( \frac{1}{N_1^2} \sum_{i: Z_i=1}  v(|\Ni_i|) + \frac{1}{N_0^2} \sum_{i: Z_i=0}  v(|\Ni_i|) \bigg)
\end{flalign*}

which completes the proof.
\end{proof}


\section{Fisher intervals  based on restricted randomizations}
\label{section:fisher-test}

\subsection{Inferential procedure}

This section simply combines together classic results on inverting Fisher tests \citep{rosenbaum2002covariance} and on exact tests with restricted randomization \citep{morgan2012rerandomization}. The inferential procedure assumes that we have imposed balance, so $Z\in\mathcal{Z}^b$, which is the case considered in the article. We begin by  describing how to obtain a p-value for the following type of sharp null
hypothesis,
\begin{equation*}
	H_{\tau^*}: Y_i(1) = Y_i(0) + \tau^* \quad \forall \, i,
\end{equation*}
then we  describe how to invert a sequence of such tests to obtain a confidence interval. Let $Z \sim \mathcal{R}$ be
any of the restricted randomization schemes we proposed that imposes exact balance on the size of the treatment groups.
A p-value for $H_{\tau^*}$ is obtained as follows:
\begin{enumerate}
	\item let $T^{obs} = \hat{\tau}(Z^{obs})$. 
	\item Define the following potential outcomes:
	\begin{equation*}
		Y^*_i(1) = \begin{cases}
			Y_i(Z_i^{obs}) &\mbox{ if} \,\,  Z_i^{obs} = 1 \\
			Y_i(Z_i^{obs}) + \tau^* &\mbox{otherwise}
		\end{cases}
		\quad and \quad
		Y^*_i(0) = \begin{cases}
			Y_i(Z_i^{obs}) &\mbox{ if} \,\,  Z_i^{obs} = 0 \\
			Y_i(Z_i^{obs}) - \tau^* &\mbox{otherwise}
		\end{cases}
	\end{equation*}
	\item For $k = 1, \ldots, M$, let $Z_k \sim \mathcal{R}$, compute $T_k = \hat{\tau}(Z_k, \Y^*(0), \Y^*(1))$.
	\item Compute the two-sided p-value $p^{(M)}_{\tau^*} = \frac{1}{M} \sum_k^M \iv(|T_k| > |T^{obs}|)$ 
\end{enumerate}
The p-value $p^{(M)}_{\tau^*}$ thus obtained is a monte-carlo approximation of the true p-value $p_{\tau^*}$ for
the test $H_{\tau^*}$. It can be made arbitrarily precise by increasing $M$. The difference with the traditional
Fisher test here occurs in Step 3, when we sample $Z \sim \mathcal{R}$, where $\mathcal{R}$ is a 
restricted randomization design. This is usually performed using rerandomization as described in the main
article.

Confidence intervals are then obtained
by inverting a sequence of such Fisher exact tests. Specifically, let $\tau^{min} < \tau^{max}$ be such that 
$p^M_{\tau^{min}} < \alpha$, $p^M_{\tau^{max}} < \alpha$, and such that there exists $\tau^{min} < \tau < \tau^{max}$ such that $p^M_{\tau} > \alpha$. These can always be found, by construction.
Let $\delta_K = (\tau^{max} - \tau^{min})/K$. A $100\times(1-\alpha)$ interval for $\tau$ can be constructed as follows:
\begin{enumerate}
	\item For $k = 1, \ldots, K$, compute $p^{M}_{\tau^{min} + k*\delta_K}$.
	\item Define:
	\begin{equation*}
		k^{(low)} = \min \{ k: p^{M}_{\tau^{min} + k*\delta_K} > \alpha \} \quad \mbox{ and } \quad k^{(high)} = \max \{ k: p^{M}_{\tau^{min} + k*\delta_K} > \alpha \} 
	\end{equation*}
	\item The interval $[ \tau^{min} + k^{(low)} \cdot \delta_K, \tau^{min} + k^{(high)} \cdot \delta_K ]$ is a $100\times(1-\alpha)$ interval for $\tau$
\end{enumerate}

\subsection{Simulation study}

This simulation compares the size of the Fisher intervals obtained with the balanced optimal 
model-assisted design, to that of the Fisher intervals obtained with balanced complete randomization. The 
setup is as follows. We generated 100 Erdos-Renyi graphs, with $N$ nodes, and parameter $p=0.15$. 
For each graph, we computed the mean square error for the associated normal sum model with 
parameters $\mu = 1$, $\sigma=2$, $\gamma=1$, and $\tau = 1$. Then for each graph we generated
200 independent realizations of the potential outcomes vector from the true model, and for each
realization we computed the Fisher intervals from balanced complete randomization, and from balanced
optimal randomization with $\alpha = 0.05$. Thus in total, we obtained 20000 intervals for each method. We 
then then repeated the same step but with misspecified models. Specifically, we consider the 'small'
misspecification case in which we randomly modify $5\%$ of the edges in each graph, and the 'large'
misspecification case in which we randomly modify $10\%$ of the edges in each graph. In each
misspecification case, we then proceed as if the misspecified graph was the correct graph (so we compute
the mean square error based on the misspecified graph), but we generate the potential outcomes from the true model. Then
as in the correctly specified case, we compute the size of the Fisher intervals obtained using our method
to those obtained under balanced randomization.\\

Figures~\ref{fig:fisher} summarizes the results. Each panel shows the percentage reduction in the size
of confidence intervals obtained by using our method, for a single network (so 200 values). Each row
correspond to a different degree of misspecification. The columns indicate how we chose the network
displayed in the panels: for the panel in the first column and first row, we chose the network for which the
mean percentage reduction was smallest, in the correctly specified case. The other panels can be
interpreted similarly.\\

The key message of this plot is that our method leads to smaller confidence intervals, even under strong
misspecification, and that this reduction is consistent across different networks.

\begin{figure}[t!]
	\center
	\includegraphics[scale=0.65]{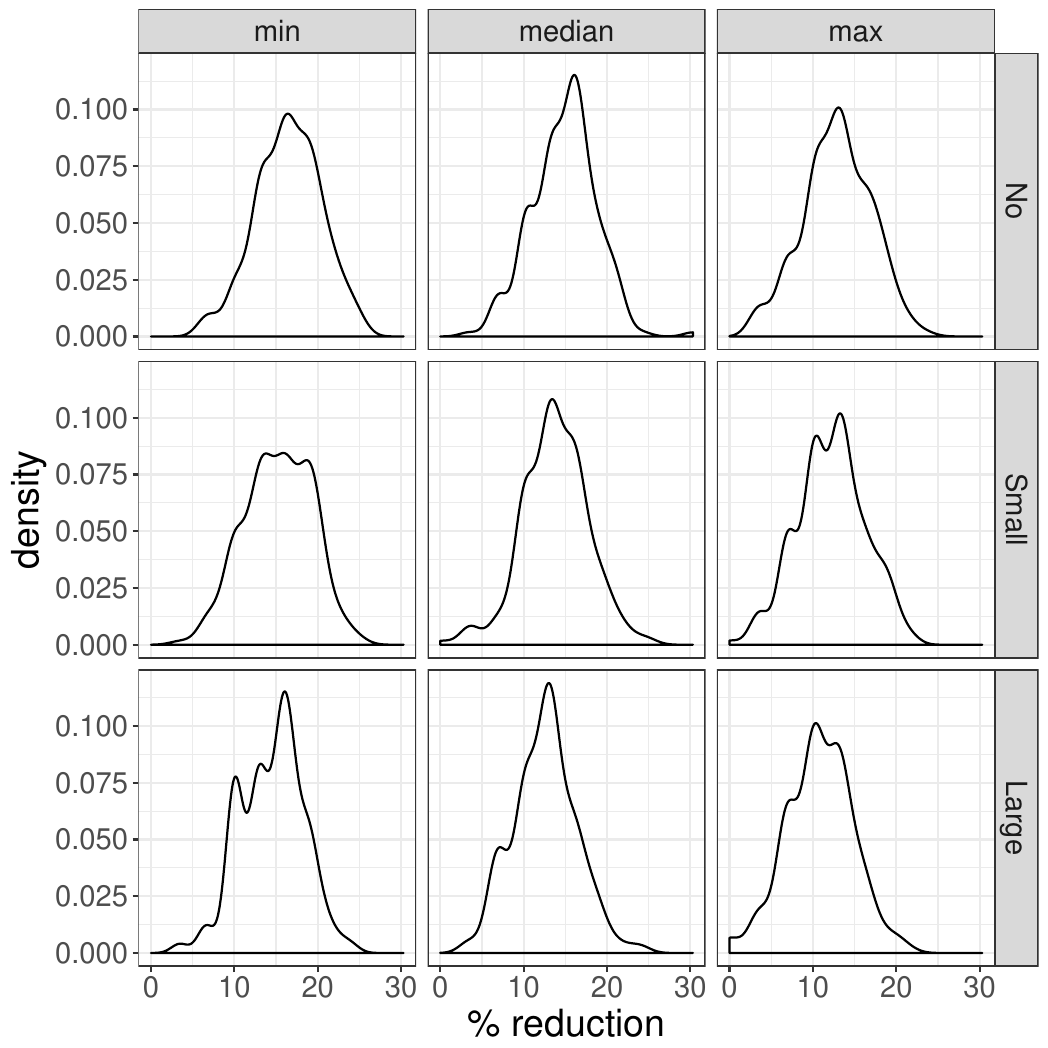}
	\caption{Reduction is the size of Fisher intervals obtained using our balanced optimal restricted 
	randomization strategy (with $\alpha = 0.05$) compared to the size of Fisher intervals obtained
	using a balanced randomization design.}
	\label{fig:fisher}
\end{figure}


\section{Numerical results and robustness to misspecification}
\label{section:results}

In this section, we report simulation results to assess the performance of the proposed randomization and re-randomization strategies against standard completely randomized allocation, Bernoulli allocation, and more recent re-randomization strategies based on these strategies.
We perform three sets of simulations.
 In Section \ref{section:res-performance}, the proposed randomization strategies are obtained by relying on diffuse prior distributions for key parameters centered around the true values.
 In Section \ref{section:res-robustness-net}, we explore comparative performance when the actual model (namely, the network used to specify the model) is misspecified.
 In Section \ref{section:res-robustness-prior}, we explore comparative performance when the prior distributions informing the proposed strategies are increasingly misspecified.


\subsection{Design of simulation experiments}
\label{section:res-setup}


We consider four families of networks:
 Erd\"os-Renyi, power law, stochastic blockmodel, and small world on a ring lattice  \citep{goldenberg2010survey}.
 We do this for convenience, but without loss of generality, since the formulas for the mean square error in Sections \ref{sec:cond-mse}--\ref{sec:marg-mse} and the theory and methods in Section \ref{sec:tm} depend on observed network statistics.
 We generate 100 networks, each with 500 nodes, from these families.
 These networks all have comparable densities ($0.08 \pm 0.02$) by design. 
The outcomes are generated according to the model in Equations \ref{eq:nsx}--\ref{eq:nsy1}, with parameters $\mu =1$, $\sigma = 2$ and $\gamma=1$.
We note that several allocation strategies described in Section~\ref{sec:marr} require solving optimization problems, for which we can only provide approximate solutions. All optimizations are carried out via stochastic optimization \citep{goldberg1988genetic}. We discuss the variability in the results due to this approximation when appropriate.


\subsection{Comparative performance analysis}
\label{section:res-performance}

The goal of this set of simulations is to quantify the order of magnitude of improvements in integrated mean squared error an analyst can expect, under controlled conditions. 
In these simulations, we compare the performance of the different estimators when the data are simulated from the model in Equations \ref{eq:nsx}--\ref{eq:nsy1}. 
For each of the 400 networks described in Section~\ref{section:res-setup}, we generate 300 assignments for each of the methods described in Section~\ref{sec:tm}. For each assignment we compute the mean square error in Equation \ref{eq:rerand}.
Thus the results here compare performance of the randomization strategies coupled with the simple difference-in-means estimator. We postpone the discussion of the maximum likelihood estimator to the following section. 

\begin{figure}[t!]
  \centering
	\includegraphics[width=0.48\textwidth]{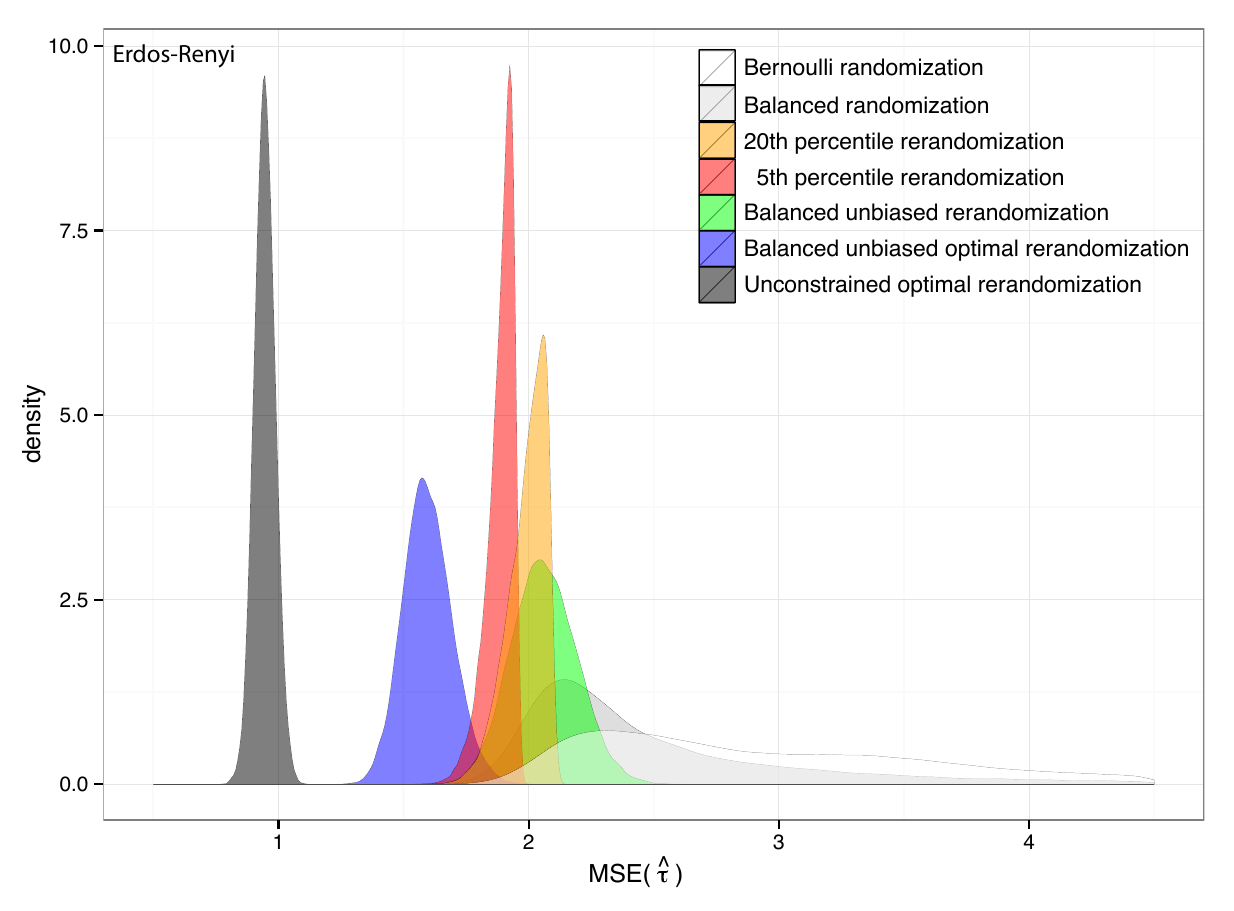}
	\includegraphics[width=0.48\textwidth]{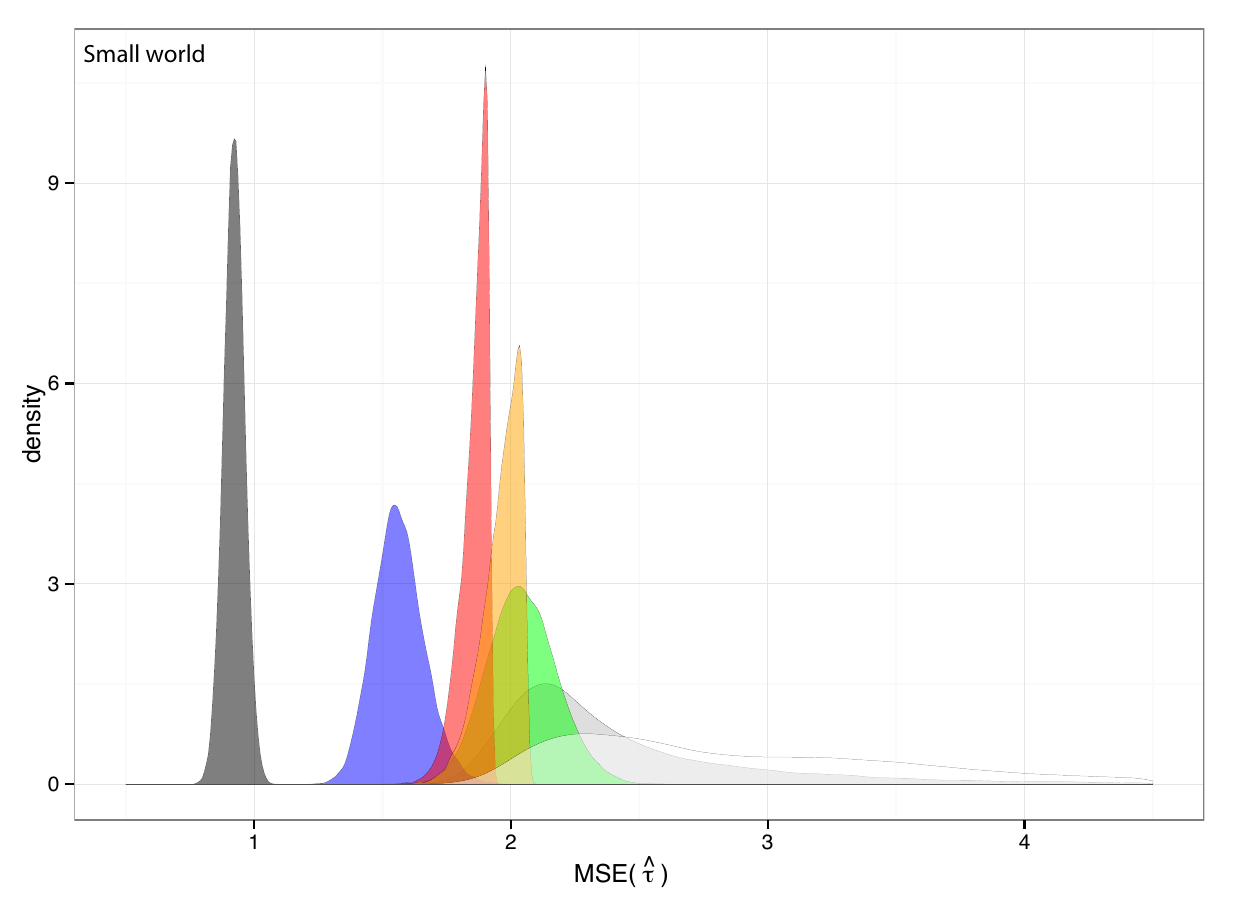}\\
  \centering
  	\includegraphics[width=0.48\textwidth]{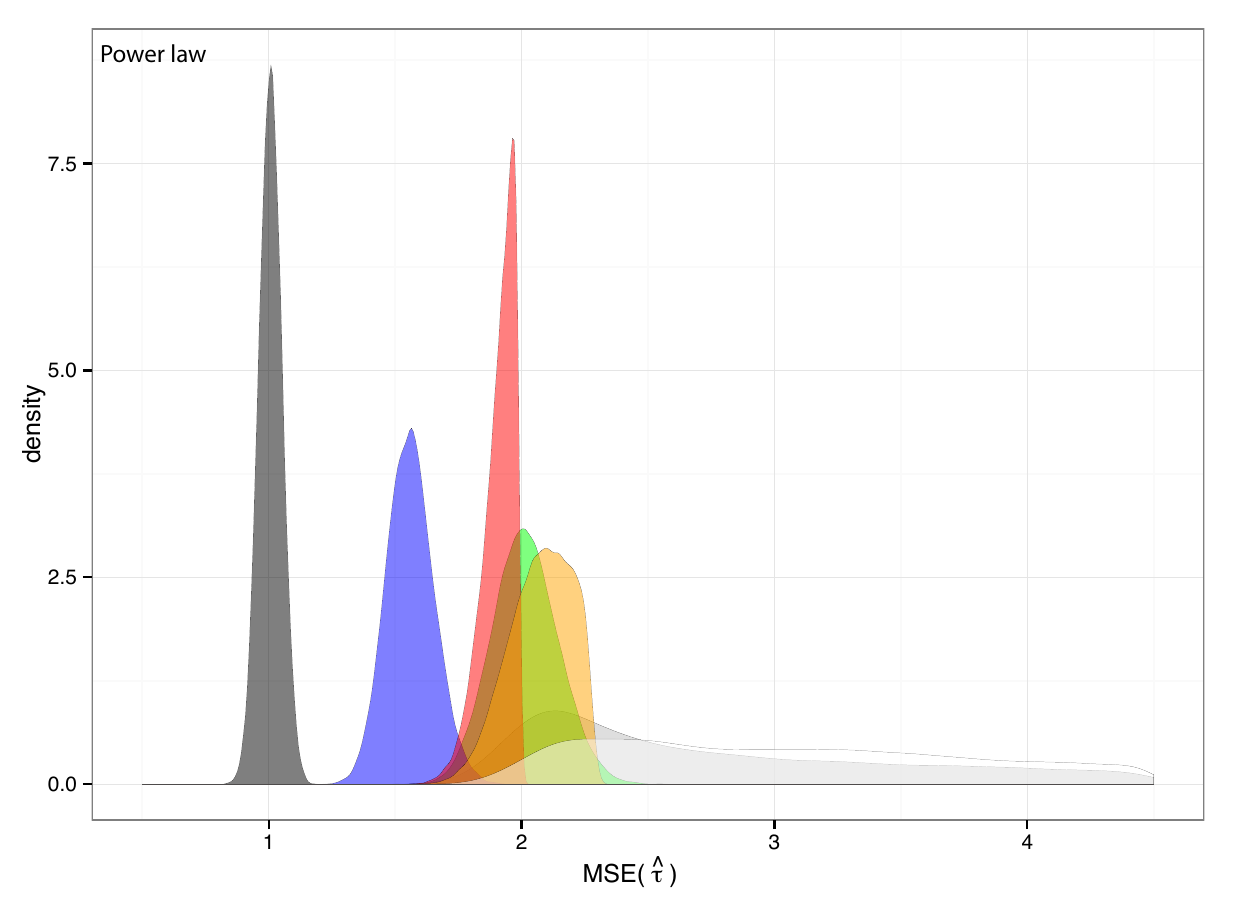}
	\includegraphics[width=0.48\textwidth]{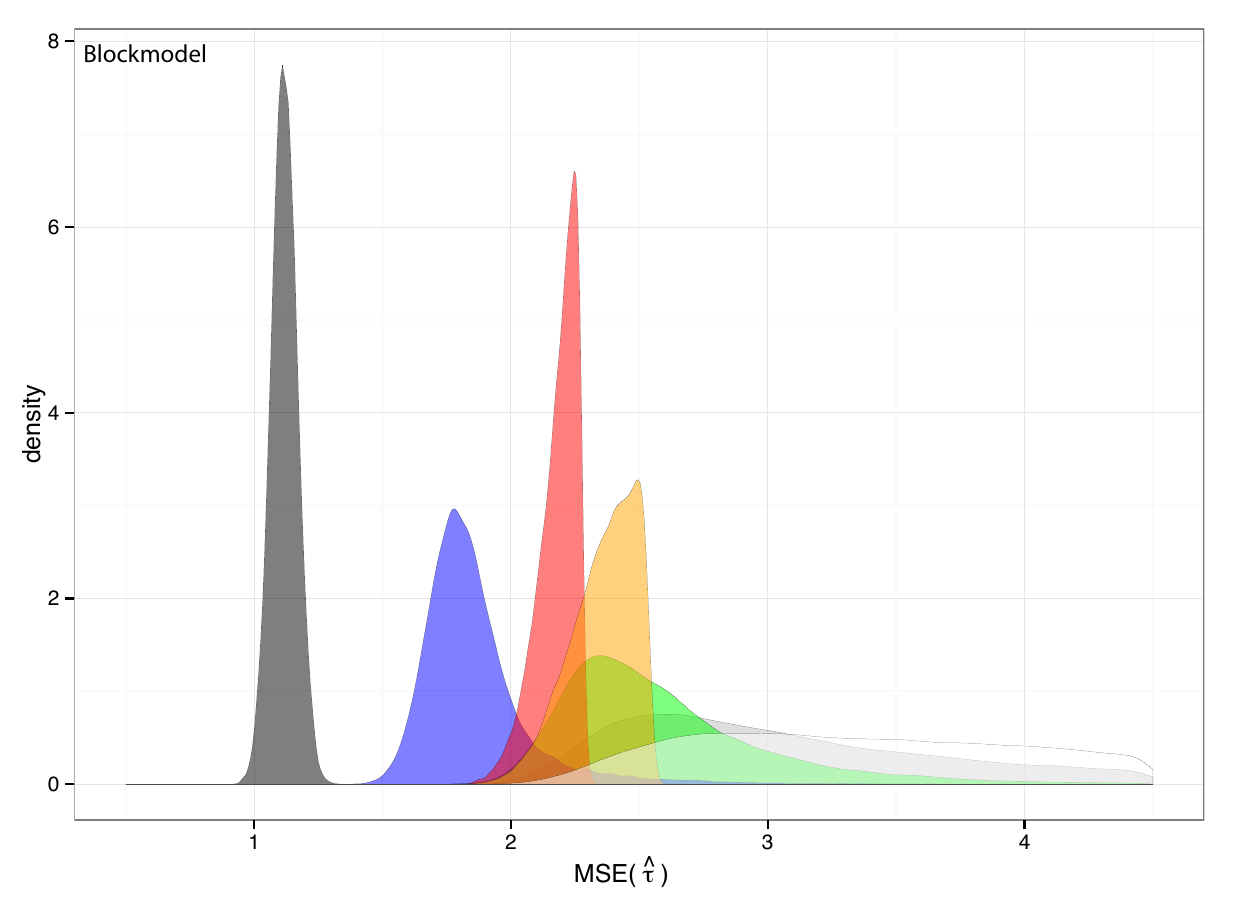}
	\caption{Distributions of the marginal MSE (Equation \ref{eq:rerand}) for seven randomization strategies, each estimated from 30,000 replicated experiments for each network family.}
	\label{fig:perf-all}
\end{figure}

\begin{figure}[b!]
  \centering
	\includegraphics[width=0.48\textwidth]{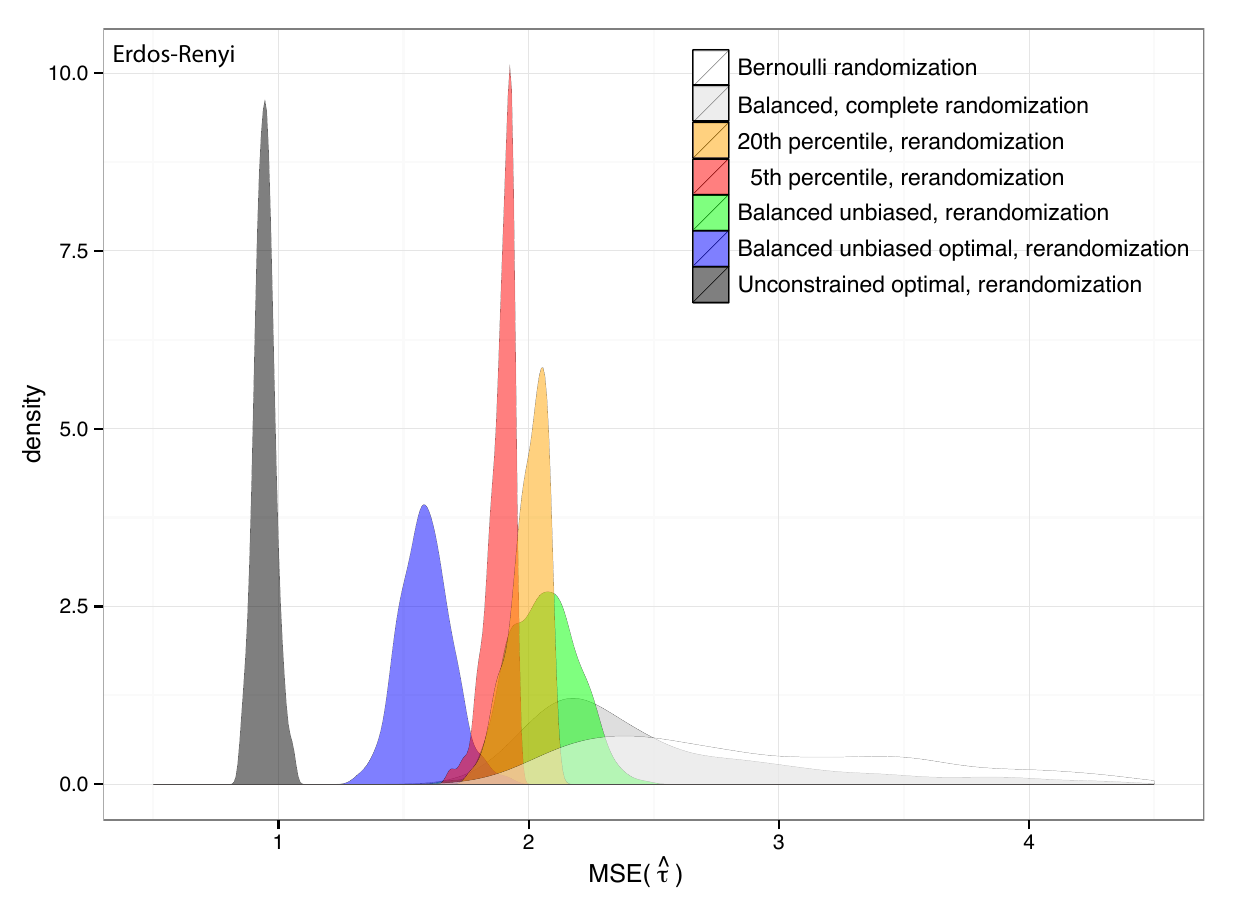}
	\includegraphics[width=0.48\textwidth]{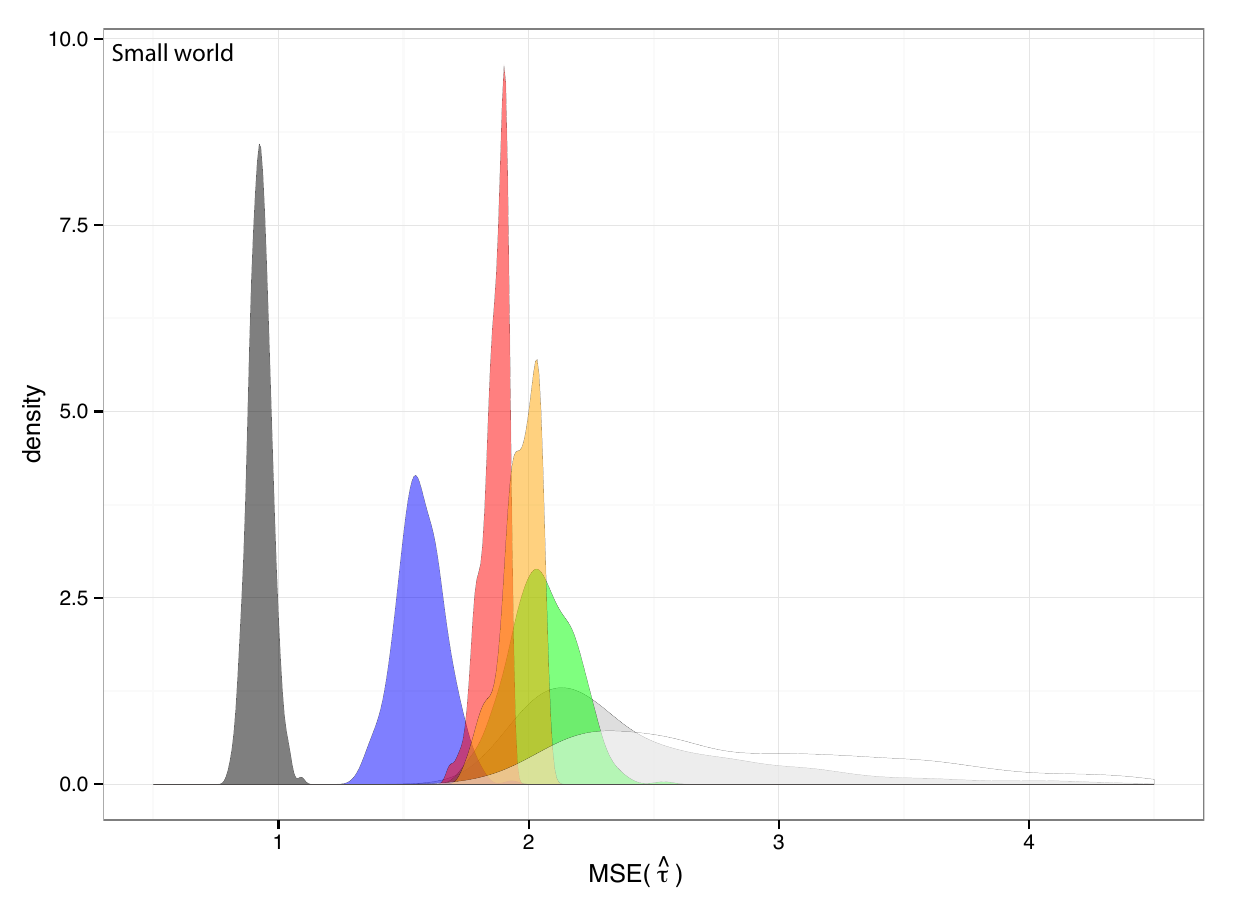}\\
  \centering
	\includegraphics[width=0.48\textwidth]{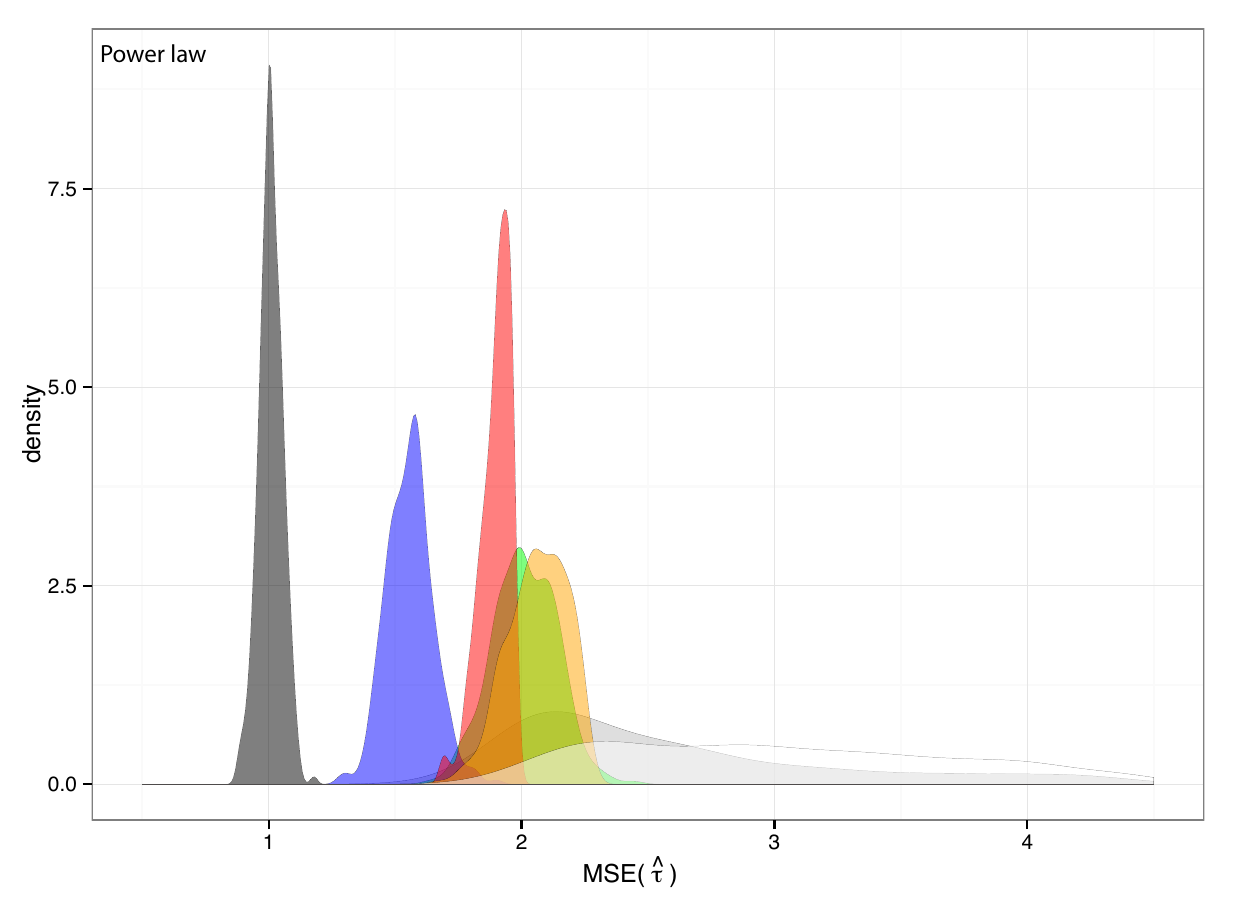}
	\includegraphics[width=0.48\textwidth]{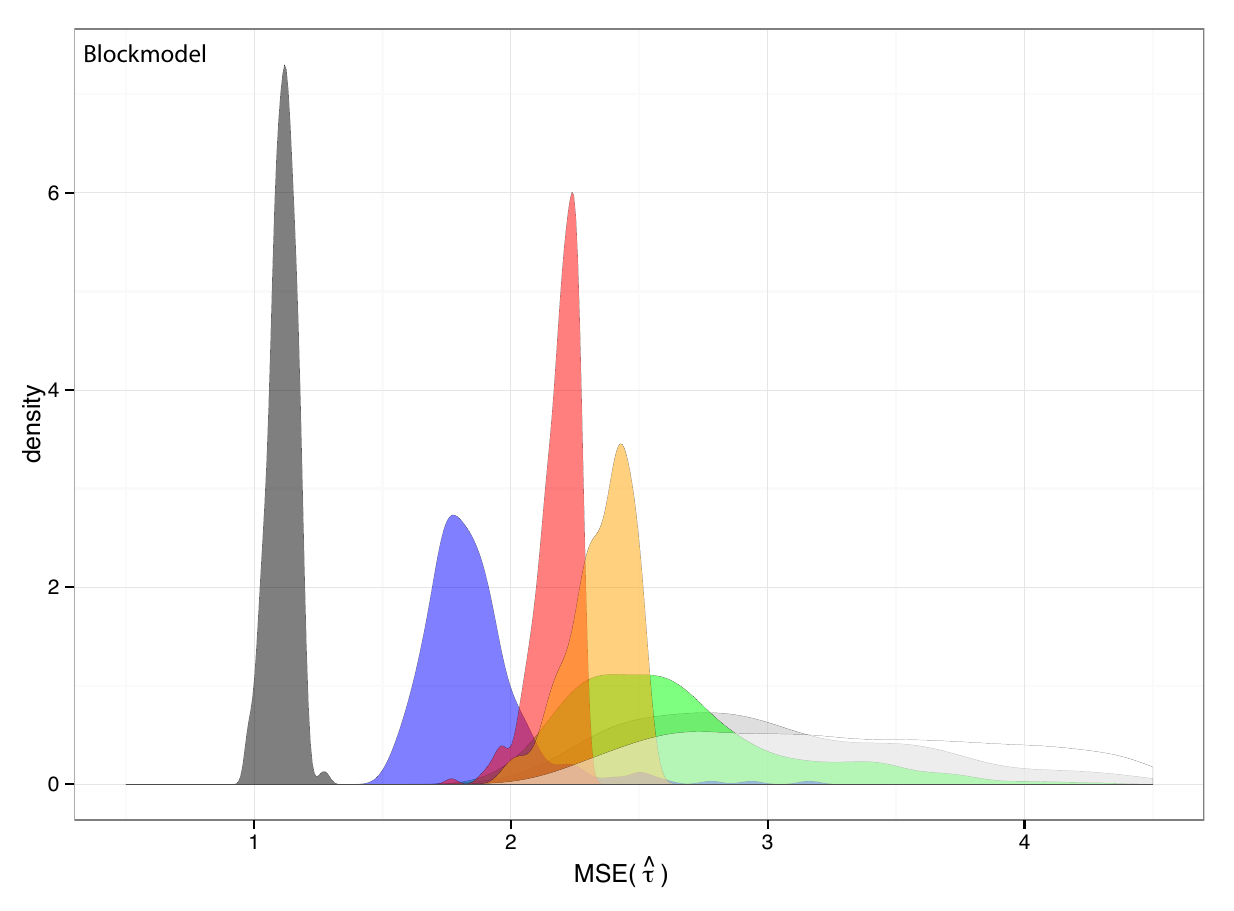}
	\caption{Distributions of the marginal mean square error (Equation \ref{eq:rerand}) for seven randomization strategies, each estimated from 300 replicated experiments on a single simulated network for each  family.}
	\label{fig:perf-individual-1}
\end{figure}

Figure~\ref{fig:perf-all} shows the mean square error densities for seven randomization strategies, estimated from 30,000 replicated experiments for each network family.
Figure~\ref{fig:perf-individual-1} shows the mean square error densities for seven randomization strategies, estimated from 300 replicated experiments for the first simulated network in each family.
In both Figures, we truncated the X axis at 5, however the mean square errors for the Bernoulli and balanced randomizations take values as high as 10.

The results suggest a number of observations.
Balancing the number of treated and control units generally improves the mean squared error over Bernoulli randomizations.
Removing bias through rerandomization, by balancing the average degree of treated and untreated units (see Equation \ref{interpretation:bias}), generally improves the mean squared error over balanced randomizations. 
Rerandomizations that effectively discard balanced randomizations with high mean square error often outperform balanced unbiased rerandomizations, with the exception of power law networks. In these networks, the balanced unbiased rerandomization still outperforms the rerandomization that keeps only allocation vectors with the 20\% highest mean square error. This happens because the degree distribution in power law networks is very skewed and even the top 20\% allocations in terms of mean square error display quite a lack of balance in terms of average degree between treated and non treated units.

We note that different rerandomization strategies explore the treatment allocation vectors with different criteria, thus these improvements are simply a consequence of the difficulty in exploring a vast space; we sampled 300 allocation vectors in a space that has roughly $2^{100}$ elements. In our experience in designing large experiments practitioners typically generate tens of thousands of allocation vectors, but only look closely at hundreds of them, so our simulation is realistic in this respect \citep[][and ongoing work by the authors]{kim2015aa}. 
The other three network families, on the other hand, have much more symmetric degree distributions, and so explicitly disregarding balance on average degree does not lead to heavy bias and higher mean square errors.

Interestingly, balanced unbiased rerandomizations that explicitly control the variance terms in the conditional mean square error (in Equations \ref{eq:31}--\ref{eq:33}) substantially improve the mean squared error over balanced rerandomizations based on the overall mean square error. This suggest that is is unlikely to find allocation vectors with good variance control in a small set of allocations.
Unconstrained rerandomizations that directly control the sum of bias and variance terms substantially improve the mean squared error over balanced unbiased optimal rerandomizations. This is consistent with the findings in classical estimation tasks, where a small increase in bias may lead to larger reductions in variance, and thus to lower mean square error.
Lastly, the gap between the mean square error of unconstrained optimal rerandomizations and the mean square error of balanced unbiased optimal rerandomizations depends on the family of networks we consider.

Overall, model-assisted rerandomizations perform better, under ideal conditions. The theory in Section \ref{sec:theory} provides assurances in terms of unbiasedness when the model does not hold.
%


\subsection{Robustness to network misspecification}
\label{section:res-robustness-net}

The goal of this set of simulations is to quantify the loss in performance of the  randomization strategies we are considering when the network the model conditions on is misspecified.

We perturbed each of the 400 networks simulated for Section~\ref{section:res-setup} by randomly rewiring different proportions of the edges. For each perturbed network we generated 100 allocation vectors using six randomization strategies---those considered in the previous section with the exclusion of the Bernoulli randomizations (same color scheme as above). Perturbations in the network only affect the proposed model-assisted rerandomization strategies, which rely on explicit bias and variance terms that now depend on the perturbed network, while the outcomes are generated according to a model that relies on the unperturbed network. 
In addition, we consider the randomization strategy that minimizes the analytical expression for the mean square error of the maximum likelihood estimator, in Equation \ref{eq:cond-mse_mle}, which also leverages the perturbed network for estimating the ATE (in purple).
We evaluated assignments in terms of marginal mean squared error, computed using the unperturbed networks. 

Figure~\ref{fig:robustness_net} displays the resulting mean square errors (mean $\pm$ 2 standard deviations) for the seven randomization strategies described above, and for the MLE under a balanced complete randomization as a baseline for the MLE (in pink). This baseline allows us to quantify the effects model misspecification on the mean square error because of failures in the estimation task only (when treatment is assigned using balanced complete randomizations), and to contrast it with the effects model misspecification on the mean square error because of failures in both estimation and optimal treatment allocation tasks.
The four panels in Figure~\ref{fig:robustness_net} show the mean square error for the four different network families. The X axis measures the fraction of edges rewired that defines the severity of the network perturbation. For instance, at 0.01 we rewire 1\% of the edges; in networks with 500 nodes and density 0.15, this corresponds to 188 edges on average.
At zero, mean square errors correspond to unperturbed networks. 
\begin{figure}[t!]
 \centering
  \includegraphics[width=0.48\textwidth]{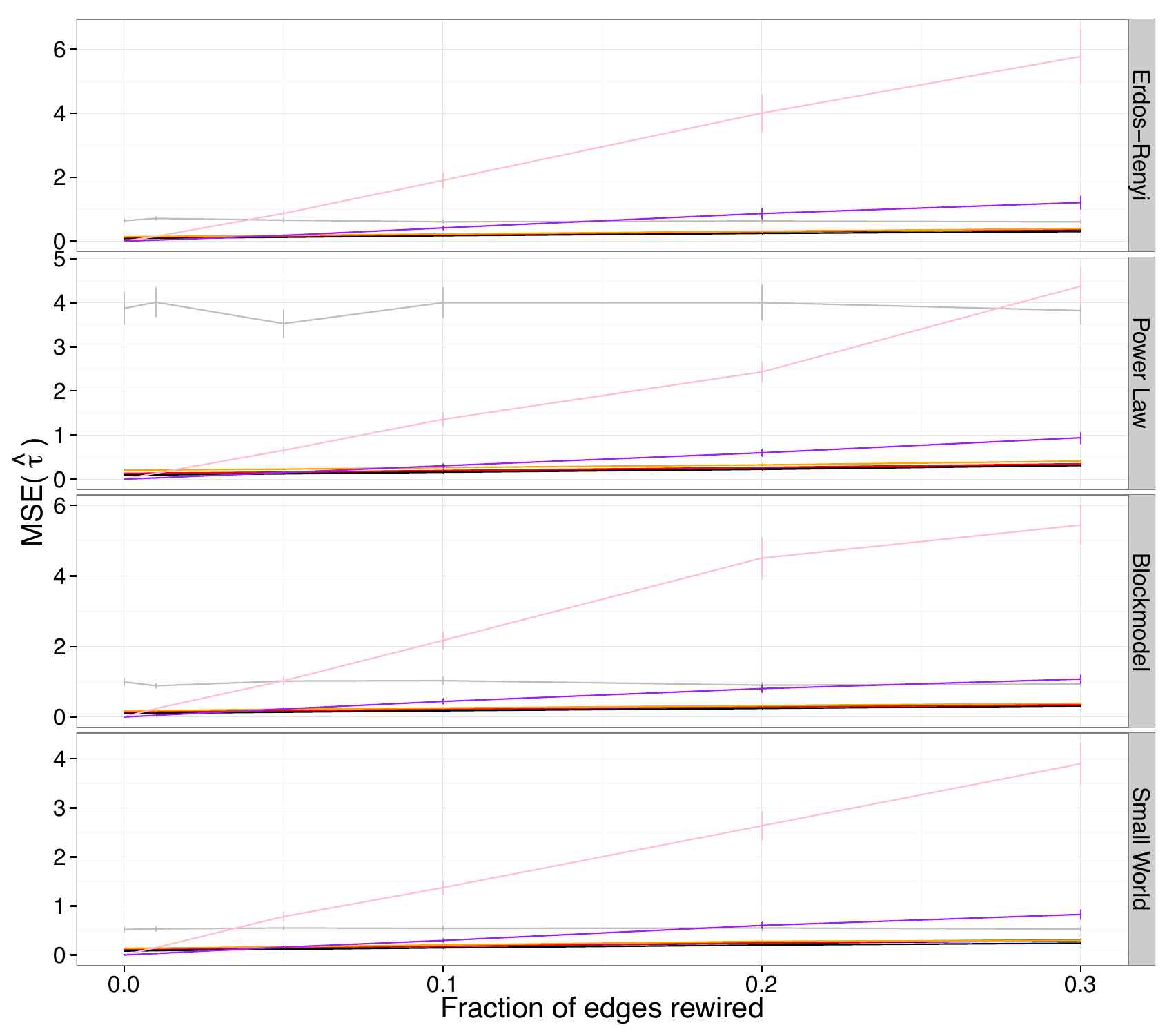}
  \includegraphics[width=0.51\textwidth]{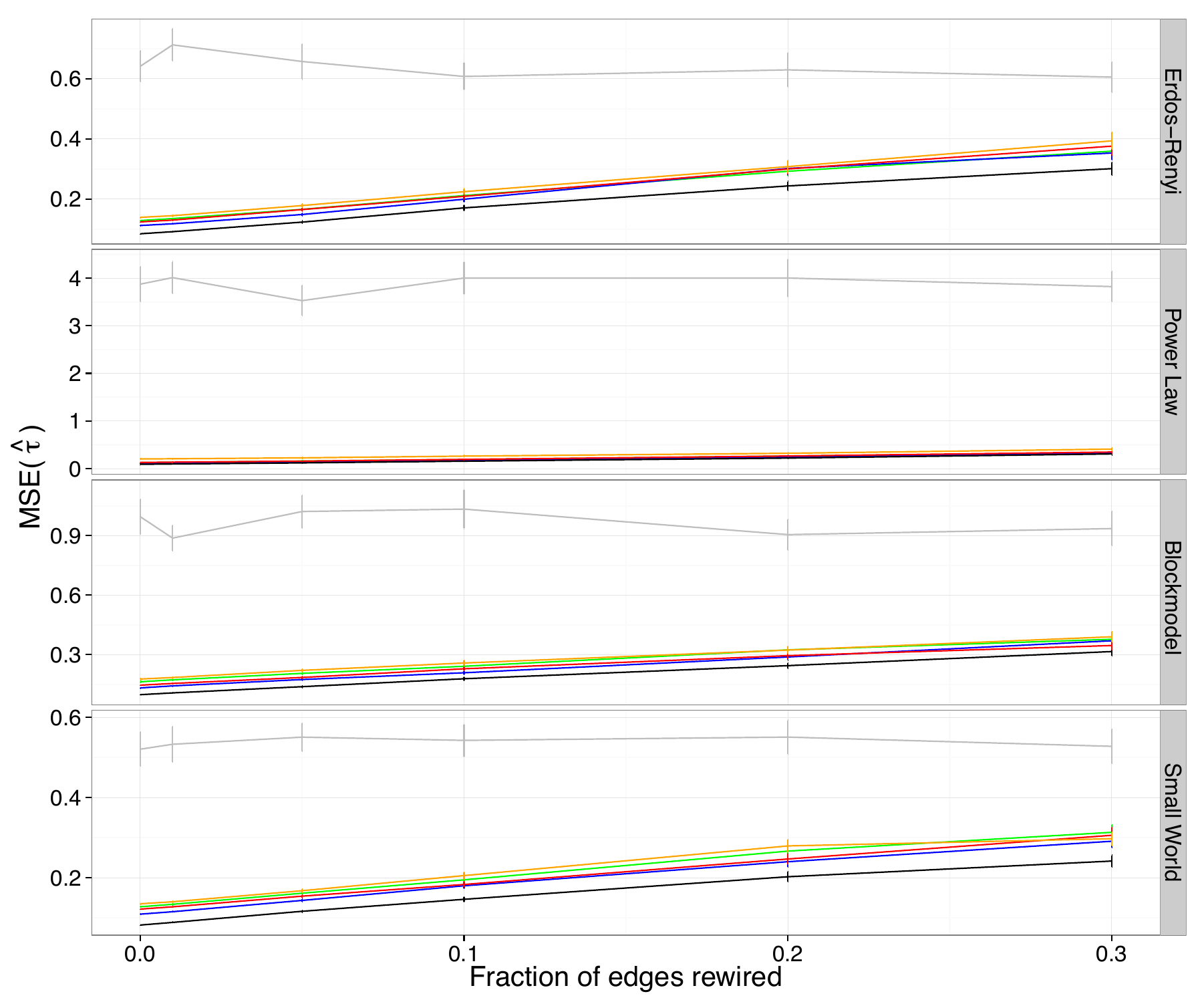}
  \caption{Left: Robustness of eight randomization strategies (colors as in previous Figures, MLE optimal in purple, MLE baseline in pink) to increasing perturbations of the network underlying the model for the outcomes in Equations \ref{eq:nsx}--\ref{eq:nsy1}. Right: Focus on the robustness to model misspecification of the six randomization strategies for the difference-in-means estimator.}
  \label{fig:robustness_net}
\end{figure}

The results suggest a few observations. 
There is a clear contrast between randomization strategies based on the MLE and those based on the difference-in-means estimator.
While strategies targeting the MLE outperform the other strategies in the absence of model misspefication (i.e., no edges rewired), even for modest misspecification (i.e., 5\% edges rewired) their mean square error increases substantially and exceeds that of strategies based on the  difference-in-means estimator.
Perhaps surprisingly, the balanced complete randomization for the MLE (pink curve) performs worst than the optimizing treatment assignment for the MLE (purple curve)
for the range of misspecification explored. 
%
This over-sensitivity to model misspecification makes MLE-based randomization strategies, and MLE estimation of the ATE, unattractive options in practice.

In contrast, randomization strategies based on the difference-in-means estimator are generally insensitive to increasing amounts of misspecification, which is plausible since this estimator does not depend on the network. 
Any mount of misspecification (in the range we consider) does not alter the ordering the proposed rerandomization strategies suggested in Section \ref{section:res-performance}, in terms of average marginal mean square error over the simulated networks. 


\subsection{Robustness to prior misspecification}
\label{section:res-robustness-prior}

The goal of this set of simulations is to quantify the loss in performance of the randomization strategies we are considering when parameters in the model for the outcomes are misspecified.

\begin{figure}[b!]
\begin{minipage}{0.45\textwidth}
 \center
 \vspace{-25pt}
  \begin{tabular}{c|ccc}
		Set no. &$\mu$ & $\gamma$ & $\sigma$ \\ \hline
		1 & 1  &   1  &  2\\
		2 &20 & 20 & 0.1 \\		
		3 & 0.1 & 0.1 & 0.1 \\
		4&20 & 20 & 20\\		
		5 & 0.1 & 20 & 0.1 \\	
		6&0.1 & 20 & 20 \\
		7&0.1 & 0.1 & 20 \\			
		8 & 20  & 0.1 & 0.1 \\
		9&20 & 0.1 & 20 \\
	\end{tabular}
	\vspace{20pt}
	\renewcommand{\figurename}{Tab.}
	\setcounter{figure}{0}
	\caption{Parameters values for Figure~\ref{fig:robustness_prior_all}.}
	\label{table:params} 
    \setcounter{figure}{3}
 \vspace{-25pt} 
\end{minipage}
\begin{minipage}{0.5\textwidth}
 \center
  \includegraphics[width=1\textwidth]{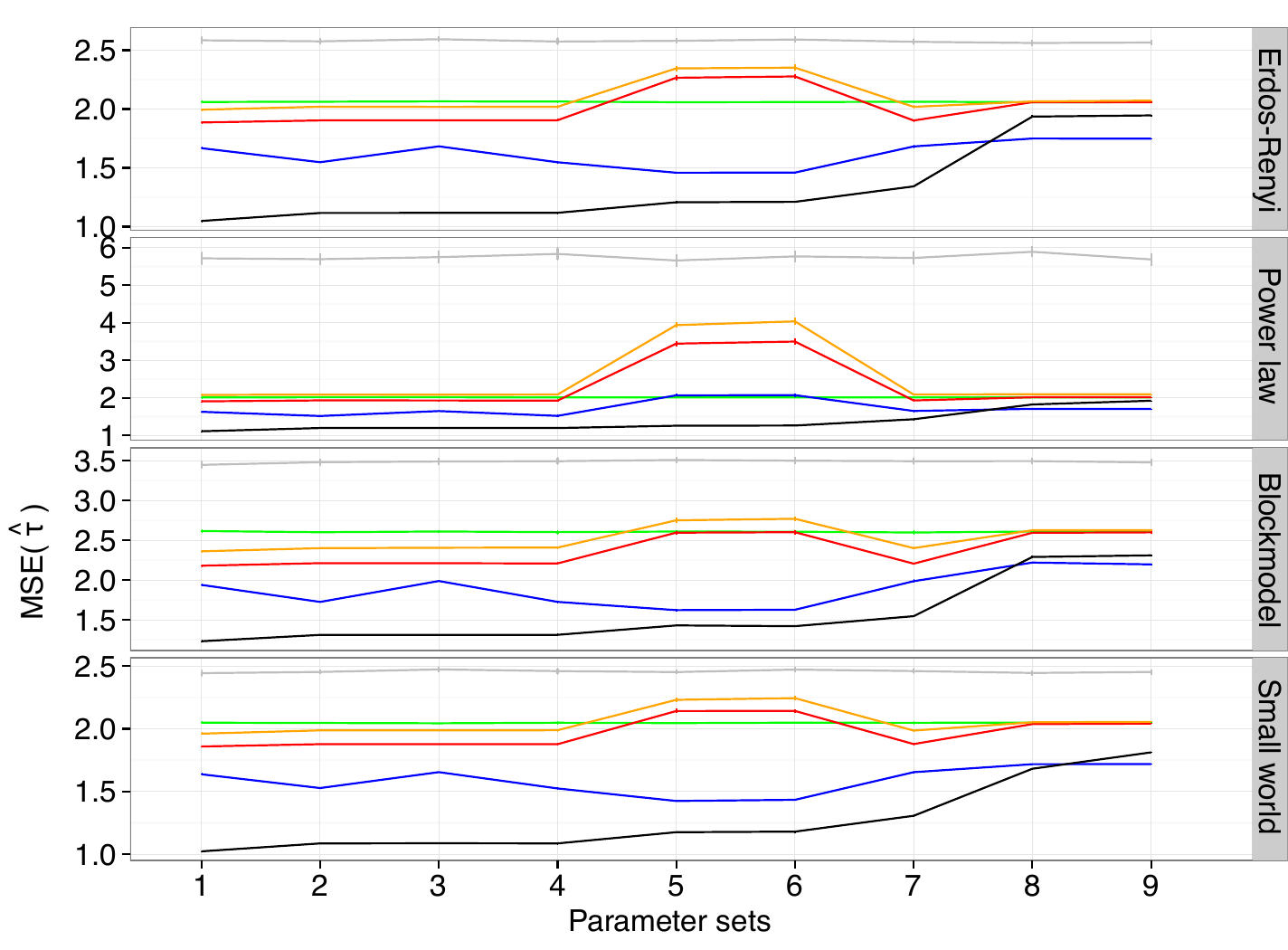}
  \caption{Robustness to prior misspecifications.}
  \label{fig:robustness_prior_all}
 \vspace{-10pt} 
\end{minipage}
\end{figure}

For each of the 400 networks used in the previous simulations, we generated 100 assignments from each of the six randomization strategies based on the difference-in-means estimator, for each of the nine sets of parameters specified in Table~\ref{table:params}. 
Recall that $\gamma^2$ is the variance of the outcomes, and $(\mu,\sigma^2)$ are mean and variance of the individual features that induce correlation among the outcomes along a given network.
The parameter sets are listed in increasing order or average marginal mean square error expected under unconstrained optimal rerandomizations. The first set of parameter values gives the values used to generate the outcomes.

Figure~\ref{fig:robustness_prior_all} shows the resulting mean sqartuare errors (mean $\pm$ 2 standard deviations) for the six randomization strategies (same color scheme as above). 
The results suggest a few observations. 
The balanced complete randomization strategy (in gray) is insensitive to the changes in the model since it does not rely on any aspect of it for assigning treatment. 
The two balanced rerandomization strategies (5\% in red, 20\% in orange) select allocations based on their conditional mean square error, which is computed using misspecified parameters; these strategies suffer in settings where both the parameter that controls the bias $\mu$ is wrongly assumed to be negligible and the parameter that controls the variability in the outcomes $\gamma$ is wrongly assumed to be much bigger than its real value. 
The balanced unbiased rerandomization strategy (in green) is insensitive to misspecification; it disregards the variance components of the mean square error, thus eliminating sensitivity to the misspecification of $\gamma$ and $\sigma$, and it selects allocations that zero out the term $\delta_{\mathcal{N}}$ in Equation \ref{interpretation:bias}, thus eliminating any potential adverse effects due to the misspecification of $\mu$.
The balanced unbiased optimal rerandomization strategy (in blue) is generally robust to parameter misspecification, while achieving low mean square error. 
Interestingly, the optimal unconstrained rerandomization strategy (in black), which despite parameter misspecifications achieves the lowest mean square error, in settings 8--9 suffers from trading too little bias (which equals $\mu\cdot\delta_{\mathcal{N}}$) for variance, since it wrongly assumes a high value for $\mu$, and thus looses its advantage over the balanced unbiased optimal rerandomization strategy.

\section{Analytical derivations}
\label{sec:derivations}

Notes on the appendix. As in the main text, $A$ will denote the \textit{extended adjacency matrix} of 
the graph. That is, it is the adjacency matrix with all diagonal terms equal to 1. 
For clarity, we will distinguish all expectations, variances and covariances with respect to the randomization distribution, denoted by $\sexpz,\svarz,\scovz$, from the expectations, variances and covariances with respect to the model, denoted by $\sexpy,\svary,\scovy$. Expectations, variances and covariances without subscripts are to be understood as joint operations over the randomization distribution and the model.


\subsection{Model for the observed data for the normal-sum model}
\label{sec:yobs}

We first derive the correlation between control potential outcomes
\begin{equation}
\label{eq:marg:cov}
\scovy[Y_i(0), Y_j(0)] = \begin{cases}
	|\Ni_i \cap \Ni_j|\sigma^2, &\text{if $i \neq j$}\\
	|\Ni_i \cap \Ni_j|\sigma^2 + \gamma^2, &\text{if $i=j$}.
\end{cases}
\end{equation}
It follows that
\begin{flalign*}
	\scovy[\Yw, \Yw | \W] &= \scovy[\Yzero + \W \tau, \Yzero + \W \tau | \W] \\
	&= \scovy[\Yzero, \Yzero] \\
	&= A'A \sigma^2 + I \gamma^2.
\end{flalign*}

It then follows that the observed model can be written

\begin{equation}\label{eq:obs-model}
	\Y^{\rm obs} = \Y(\W^{\rm obs}) | \W^{\rm obs} \sim \mbox{Multivariate-Normal}(A \mu + \tau \W^{obs}, \gamma^2 I + \sigma^2 A A^t).
\end{equation}
%


\subsection{Derivation of the conditional mean square error for the normal-sum model}
\label{sec:deriv-ns}

It follows from the calculations in Appendix~\ref{sec:yobs} that:
\begin{flalign*}
	\svary[\hat{\tau} | \W] &= \svary[\bm{\omega}(\W)' \Yzero] \\
	&= \bm{\omega}'(\W) \scovy[\Yzero, \Yzero] \bm{\omega}(\W)\\
	&=\bm{\omega}'(\W) \bigg(A'A \sigma^2 + I \gamma^2\bigg) \bm{\omega}(\W)
\end{flalign*}
where $\bm{\omega}(\W) = \W /  N_1 - (1-\W)/ N_0$. So we have shown that:
\begin{equation}
\textstyle
	\svary[\hat{\tau} | \W] = \bm{\omega}'(\W) \bigg(A'A \sigma^2 + I \gamma^2\bigg) \bm{\omega}(\W).
\end{equation}
It is quick to derive that:
\begin{flalign*}
	\sexpy[\hat{\tau} | \W] &=  \sexpy\bigg[\frac{1}{N_1} \sum_i Z_i Y_i(1) - \frac{1}{N_0} \sum_i (1-Z_i) Y_i(0)  | \W \bigg]\\
	&=\tau + \sum_i \bigg(\frac{Z_i}{N_1} - \frac{(1-Z_i)}{N_0}\bigg) \sexpy[Y_i(0)] \\
	&= \tau + \sum_i \bigg(\frac{Z_i}{N_1} - \frac{(1-Z_i)}{N_0}\bigg) \sexpy[\sum_{j\in \Ni_i} X_j]\\
	&= \tau + \sum_i \bigg(\frac{Z_i}{N_1} - \frac{(1-Z_i)}{N_0}\bigg) |\Ni_i| \mu\\
	&= \tau + \delta_\Ni \mu
\end{flalign*}
where $\delta_\Ni = \frac{\sum_{Z_i=1} \Ni_i}{N_1} - \frac{\sum_{Z_i=0} \Ni_i}{N_0}$. Thus, we immediately have:

\begin{equation}
	\mbox{Bias}(\hat{\tau}| \W)^2 \equiv  \sexpy[\hat{\tau} - \tau | \W]^2 = \mu^2 (\delta_{\mathcal{N}})^2,
\end{equation}
which gives the following conditional mean square error:
\begin{equation}
\textstyle
	\mbox{MSE}(\hat{\tau} | \W) = \mu^2 (\delta_{\mathcal{N}})^2 +\bm{\omega}'(\W) \bigg(A'A \sigma^2 + I \gamma^2\bigg) \bm{\omega}(\W). \nonumber
\end{equation}


\subsection{Derivation of the marginal mean square error for the normal-sum model}
\label{sec:mse-all-analysis}

The purpose of this section is to derive the analytical expression for:
\begin{equation*}
\sexp[(\hat{\tau} - \tau)^2] = \sexpy[\sexpz[(\hat{\tau}-\tau)^2 | \Y]].
\end{equation*}
We start by noticing that:
\begin{equation}
	\sexp[\hat{\tau}] = \sexpy[\sexpz[\hat{\tau}] | \Y ] = \sexpy[\tau] = \tau 
\end{equation}	
so $\mbox{Bias}(\hat{\tau} ) = 0$. Also, we have:

\begin{flalign*}
	\svar[\hat{\tau}] &= \sexpy[\svarz[\hat{\tau} | \Y]] + \svary[\sexpz[\hat{\tau} | \Y]] \\
	&= \sexpy[\svarz[\hat{\tau} | \Y]] + \svary[\tau] \\
	&= \sexpy[\svarz[\hat{\tau} | \Y]] 
\end{flalign*}

So we have:

\begin{equation}
	\mbox{MSE}(\hat{\tau}) = \sexpy[\svarz[\hat{\tau} | \Y]]
\end{equation}

So we can focus on this quantity. We have

\begin{equation}
	\sexpy[\svarz[\hat{\tau} | \Y]] = \frac{\sexpy[S_1^2]}{N_1} + \frac{\sexpy[S_0^2]}{N_0}
\end{equation}

I'll start with $\sexpy[S_1^2]$ and we'll see that it's actually equal to $\sexpy[S_0^2]$. Also, I'll write $Y_i = Y_i(1)$ and $\bar{Y} = \bar{Y}(1)$ in order to simplify the notation. Now, let:

\begin{flalign*}
	D_i &= (Y_i - \bar{Y}_i)^2 \\
	      &= Y_i^2 + \bar{Y}^2  - 2Y_i \bar{Y}
\end{flalign*}

Remember that we have:

\begin{flalign*}
	\sexpy[Y_i] &= |\Ni_i| \mu + \tau\\
	\svary[Y_i] &= \gamma^2 + |\Ni| \sigma^2\\
	\scovy[Y_i, Y_j] &= |\Ni_i \cap \Ni_j| \sigma^2 \,\,\,\, i\neq j
\end{flalign*}

so:

\begin{flalign*}
	S_1^2 &= \frac{1}{N-1} \sum_i D_i \\
	&= \frac{1}{N-1} \sum_i Y_i^2 + \frac{1}{N-1} \sum_i \bar{Y}^2 - \frac{2}{N-1} \sum_i Y_i \bar{Y} \\
	&= \frac{1}{N-1} \sum_i Y_i^2 + \frac{N}{N-1} \bar{Y}^2 - \frac{2N}{N-1}  \bar{Y} ^2 \\
	&= \frac{1}{N-1}  \sum_i Y_i^2 - \frac{N}{N-1} \bar{Y}^2\\
	&= B_1 - B_2
\end{flalign*}
where  $B_1 = \frac{1}{N-1}  \sum_i Y_i^2$ and $B_2 = \frac{N}{N-1} \bar{Y}^2$. We have
\begin{flalign*}
	\sexpy[Y_i^2] &= \svary[Y_i] + \sexpy[Y_i]^2 \\
		&= \gamma^2 + |\Ni_i| \sigma^2 + (|\Ni_i| \mu + \tau)^2
\end{flalign*}
so 
\begin{equation}
	\sexpy[B_1] = \frac{1}{N-1} \left( N\gamma^2 + N\sigma^2 \overline{|\Ni|} + \sum_i ( \overline{|\Ni_i|}\mu + \tau)^2 \right)
\end{equation}
and
\begin{flalign*}
	\sexpy[\bar{Y}^2] &= \svary[\bar{Y}] + \sexpy[\bar{Y}]^2 \\
			&= \frac{1}{N^2} \textbf{1}^t \Sigma \textbf{1} + (\tau + \mu \overline{\Ni})^2
\end{flalign*}

%
%
%
%
%
Now note that:
\begin{flalign*}
	\textbf{1}^t\Sigma \textbf{1} &= N\gamma^2 + \sigma^2 \sum |\Ni_i| + 2\sum_{i<j} |\Ni_i \cap \Ni_j| \sigma^2 \\
			&= N\gamma^2 + N\sigma^2 \overline{|\Ni|} + 2\sigma^2 \sum_{i<j} |\Ni_i \cap \Ni_j| 
\end{flalign*}
So 
\begin{flalign*}
	\sexpy[B_2] &= \frac{1}{N(N-1)} \textbf{1}^t\Sigma \textbf{1} + \frac{N}{N-1} (\tau + \mu \overline{|\Ni|})^2 \\
	&= \frac{1}{N-1} \gamma^2 + \frac{1}{N-1} \sigma^2 \overline{|\Ni|} + \frac{2}{N(N-1)} \sigma^2 \sum_{i<j}|\Ni_i \cap \Ni_j| + \frac{N}{N-1} (\tau + \mu \overline{|\Ni|})^2
\end{flalign*}
and that:
\begin{flalign*}
	\sum(|\Ni_i| \mu + \tau)^2 - N(\tau + \mu \overline{|\Ni|})^2 &= N \tau^2 + 2\tau \mu N \overline{|\Ni|} + \mu^2 \sum |\Ni_i|^2 - \tau^2 N - N\mu^2 \overline{|\Ni|}^2 - 2N\tau\mu \overline{|\Ni|} \\
	&= \mu^2N (\overline{|\Ni|^2} - \overline{|\Ni|}^2)
\end{flalign*}
which leads to:
\begin{equation}
	\sexpy[S_1^2] = \gamma^2 + \sigma^2 \overline{|\Ni|} - 2 \frac{\sigma^2}{N(N-1)}\sum_{i<j} |\Ni_i \cap \Ni_j| + \mu^2 \frac{N}{N-1} (\overline{|\Ni|^2} - \overline{|\Ni|}^2)
\end{equation}

Clearly, none of the above would change for $S^2_0$ since the $\tau$'s cancel out. So finally have:

\begin{equation}
	\sexpy[\svarz[\hat{\tau} | \Y]] = (\frac{1}{N_1} + \frac{1}{N_0}) \left( \gamma^2 + \sigma^2 \overline{|\Ni|} - 2 \frac{\sigma^2}{N(N-1)}\sum_{i<j} |\Ni_i \cap \Ni_j| + \mu^2 \frac{N}{N-1} (\overline{|\Ni|^2} - \overline{|\Ni|}^2)   \right)
\end{equation}

Note: a simple sanity check is to look at what would happen if there was no network. That is, if $|\Ni_i| = 1$ for all $i$, and $|\Ni_i \cap \Ni_j| = 0$ for all $i\neq j$. The above formula then reduces to $(\frac{1}{N_0} + \frac{1}{N_1}) (\gamma^2 + \sigma^2)$, which is correct. This suggests a refactorization of the equation above:

\begin{equation}
\sexpy[\svarz[\hat{\tau} | \Y]] = V_1 + V_2
\end{equation}

where:

\begin{equation}
V_1 =  (\frac{1}{N_1} + \frac{1}{N_0}) (\gamma^2 + \sigma^2) 
\end{equation}

is the variance term in the absence of a network, and 

\begin{equation}
V_2 = (\frac{1}{N_1} + \frac{1}{N_0}) (\sigma^2 (\overline{|\Ni|} - 1)- 2 \frac{\sigma^2}{N(N-1)}\sum_{i<j} |\Ni_i \cap \Ni_j| + \mu^2 \frac{N}{N-1} (\overline{|\Ni|^2} - \overline{|\Ni|}^2) )
\end{equation}

is the variance term correction when a network structure is present. So in conclusion, we have:

\begin{equation} \label{eq:new-mse}
	\mbox{MSE}(\hat{\tau}) = V_1 + V_2
\end{equation}

which completes the proof.


\subsection{Analysis of the difference-in-means estimator under the normal-mean model}
\label{sec:nmm-analysis}

%
%

We consider the normal-means model, as an alternative:
\begin{flalign}
 \label{eq:nmx}
	X_j &\overset{iid}{\sim} \hbox{ Normal }(\mu, \sigma^2)\\
 \label{eq:nmy0}
	Y_i(0) \mid X&\overset{ind}{\sim} \hbox{ Normal }(\frac{1}{|\Ni_i|}\textstyle \sum_{j \in \Ni_i} X_j, \gamma^2) \\
 \label{eq:nmy1}
	Y_i(1) &= Y_i(0) + \tau.
\end{flalign}
It easy to verify that for all $\W$, we have: $\sexpy[\hat{\tau} | \W] = 0$. Then, as in the sum model, the variance can 
be expressed as:
\begin{equation*}
	\svary[\hat{\tau} | \W] = \omega(\W)^T \svary[\Y(\textbf{0}) | \W] \omega(\W) \\
\end{equation*}
where:
\begin{flalign*}
	\svary[\Y(\textbf{0}) | \W] &= \sexpy[\svar[\Y | X,\W]|\W] + \svary[\sexpy[\Y | X,\W]|\W]\\
	&= \sexpy[\gamma^2 I | \W]  + \svary[\tilde{A}X | \W] \\
	&= \gamma^2 I + \sigma^2 \tilde{A} \tilde{A}^T
\end{flalign*}
where $\tilde{A}$ is the matrix such that $\tilde{A}_{ij} = A_{ij}  / |\Ni_i|$. And so finally
%
\begin{equation}
\label{eq:nm-cond-mse}
\mbox{MSE}(\hat{\tau} | \W) = \gamma^2 \bm{\omega}(\W)'\bm{\omega}(\W)+ \bm{\omega}(\W)' \tilde{A}\tilde{A}^T \bm{\omega}(\W)
\end{equation}
Which we can write in longer form as:
\begin{flalign*}
	\mbox{MSE}(\hat{\tau}| \W) &= \gamma^2 (\frac{1}{N_1} + \frac{1}{N_0})  \\
	&+ \frac{\sigma^2}{N_1^2} \sum_{\{i,j:Z_i=Z_j=1\}}\frac{|\mathcal{N}_i\cap\mathcal{N}_j|}{|\mathcal{N}_i||\mathcal{N}_j|} \\
	&+  \frac{\sigma^2}{N_0^2} \sum_{\{i,j:Z_i=Z_j=0\}}\frac{|\mathcal{N}_i\cap\mathcal{N}_j|}{{|\mathcal{N}_i||\mathcal{N}_j|} } \\
	&- \frac{2\sigma^2}{N_1 N_0} \sum_{\{i,j: Z_i=1 \mbox{ and } Z_i=0\}} \frac{|\mathcal{N}_i \cap \mathcal{N}_j|}{|\mathcal{N}_i||\mathcal{N}_j|} 
\end{flalign*}
The first term penalizes, as before, imbalance in the sizes of the treatment groups. The last three terms look a lot like what 
we had with the sum-model, except that we now have weighted averages.
with more painful algebra, we can derive the marginal mean square error, and show that it is:
\begin{equation*}
	\mbox{MSE}(\hat{\tau}) = \bigg(\frac{1}{N_1} + \frac{1}{N_0}\bigg) \gamma^2 + \sigma^2 \overline{\bigg(\frac{1}{|\Ni_i|}\bigg)} -  \frac{2}{N(N-1)} \sigma^2 \sum_{i<j}\frac{|\Ni_i\cap\Ni_j|}{|\Ni_i||\Ni_j|})
\end{equation*}
The different terms of the equation can once again be used as new measures of balance that are functions of network quantities, although the interpretation is slightly 
more involved.


\subsection{Analysis of the maximum likelihood estimator under the normal-sum model}
\label{sec:mle-analysis}


The naive estimator does not make any reference to the network. This is not the case for the MLE, 
and we need to introduce a distinction between true and observed network. Let $A_0$ be the 
adjacency matrix associated with the true unobserved network, and $A$ be the adjacency matrix 
associated with the noisy observed network. The model we use will be based on the observed 
network, while the evaluation will be with respect to the true network. We have shown in the 
observed model of \eqref{eq:obs-model} that the observed outcomes are jointly multivariate 
normal. Let $\textbf{v}=A\mu + \tau Z^{obs}$ be the mean, and let $\Sigma = \gamma^2 I + \sigma^2 AA'$
be the variance. We also denote by $\Sigma_0 = \gamma^2I + \sigma^2 A_0 A_0'$ the variance based
on the true covariance matrix. Finally, define $\mu^* = A\mu$ and $\mu_0 = A_0 \mu$. With this, 
standard results show that:
\begin{flalign*}
	\frac{d}{d\tau} \log P(\Y^{obs}(\W) | \mu, \sigma, \gamma)  = 0 &\Leftrightarrow (\Yw - \mu^* - \tau \W)' \Sigma^{-1} \W = 0 \\
	&\Leftrightarrow \hat{\tau}^{mle}(\W) = \frac{(\Yw-\mu^*) \Sigma^{-1} \W}{\W' \Sigma^{-1} \W}
\end{flalign*}

\begin{remark}
	In all the simulations, we plug the true value of $\mu$, $\sigma$, and $\gamma$ in the mle, in order to be consistent with what we assume known
	at design-time when we compare it with the other methods.
\end{remark}
but then under the true model, we have:
\begin{flalign*}
	\sexpy[\hat{\tau}^{mle} | \W] &= \frac{(\mu_0^*+ \tau \W - \mu^*)' \Sigma^{-1} \W}{\W' \Sigma^{-1} \W} \\
	&=  \frac{(\mu_0^*- \mu^*)' \Sigma^{-1} \W}{\W' \Sigma^{-1} \W} + \tau 
\end{flalign*}
and so the bias is:
\begin{equation*}
	\mbox{Bias}(\hat{\tau}^{mle} | \W) = \frac{(\mu_0^*- \mu^*)' \Sigma^{-1} \W}{\W' \Sigma^{-1} \W} 
\end{equation*}
The variance is quickly derived:
\begin{flalign*}
	\svary[\hat{\tau}^{mle} | \W]  &= \bigg( \frac{1}{(\W' \Sigma^{-1} \W)^2} \bigg) \svary[\Yw \Sigma^{-1} \W| \W] \\
	&= \bigg( \frac{1}{(\W' \Sigma^{-1} \W)^2} \bigg)(\Sigma^{-1} \W)' \svary[\Yw | \W] \Sigma^{-1} \W \\
	&= \bigg( \frac{1}{(\W' \Sigma^{-1} \W)^2} \bigg)(\Sigma^{-1} \W)' \Sigma_0 \Sigma^{-1} \W\\
	&= \frac{\W' \Sigma^{-1} \Sigma_0 \Sigma^{-1} \W}{(\W' \Sigma^{-1} \W)^2}
\end{flalign*}
and so finally:
\begin{equation}
 \label{eq:cond-mse_mle}
	\mbox{MSE}(\hat{\tau}^{mle} | \W) =  \bigg(\frac{(\mu_0^*- \mu^*)' \Sigma^{-1} \W}{\W' \Sigma^{-1} \W}\bigg)^2  +  \frac{\W' \Sigma^{-1} \Sigma_0 \Sigma^{-1} \W}{(\W' \Sigma^{-1} \W)^2}
\end{equation}


\section{Proofs}
\label{app:proofs}


\subsection{Proof of Corollary~\ref{cor:theory}}

\noindent \textbf{Statement of Corollary~\ref{cor:theory}}

\bigskip

Let $\hat{\tau}$ be the  estimator defined in \eqref{eq:naive-est}. We have,
\begin{equation*}
	\sexpy\bigm[\mathbb{V}_{\mathcal{Z}^b\cap\mathcal{Z}^o}[ \hat{\tau} \mid \Y]\bigm] \,\,\leq~ \sexpy\bigm[\mathbb{V}_{\mathcal{Z}^b}[ \hat{\tau} \mid \Y]\bigm].
\end{equation*}

\bigskip

\begin{proof}

The key intuition for the proof is that $\mathcal{Z}^b \cap \mathcal{Z}^o \subset \mathcal{Z}^b$ and that the assignments  that are in $\mathcal{Z}^b$  and not in $\mathcal{Z}^b \cap \mathcal{Z}^o$ have large model mean square error. \\

Notice that since $\mbox{Bias}_{\mathcal{Z}^b\cap\mathcal{Z}^o}(\hat{\tau} | \Y) = \mbox{Bias}_{\mathcal{Z}^b}(\hat{\tau} | \Y) = 0$, we have:

\begin{flalign*}
	\sexpy[\svar_{\zbo}[\hat{\tau} | \Y]] &=
	\sexpy\bigg[\mbox{Bias}_{\zbo}(\hat{\tau} | \Y)^2 + \svar_{\zbo}[\hat{\tau} | \Y]\bigg]\\
	&= \sexpy[\mbox{MSE}_{\zbo}(\hat{\tau} | \Y)] \\
	&= \sexpy[\sexp_{\zbo}[ (\hat{\tau} - \tau)^2 | \Y]]\\
	&= \sexp_{\zbo}[ \sexpy[ (\hat{\tau} - \tau)^2 | \W]]\\
\end{flalign*}
and similarly
\begin{flalign*}
	\sexpy[\svar_{\zb}[\hat{\tau} | \Y]] &= 
	 \sexp_{\zb}[ \sexpy[(\hat{\tau} - \tau)^2 | \W]] \\	 
	 	 &=  \sexp_{\zb}[ I(\W \in \zbo) \sexpy[(\hat{\tau} - \tau)^2 | \W] + I(\W \in \zb \backslash \zbo) \sexpy[(\hat{\tau} - \tau)^2 | \W]] \\
		 &= P(\W \in \zbo)\sexp_{\zb}[\sexpy[(\hat{\tau} - \tau)^2 | \W] | \W \in \zbo] + (1 - P(\W\in \zbo)) \sexp_{\zb}[\sexpy[(\hat{\tau} - \tau)^2 | \W] | \W \in \zb \backslash \zbo] \\
		 &= p \sexp_{\zbo}[\sexpy[(\hat{\tau} - \tau)^2 | \W] ]  + (1 - p) \sexp_{\zb}[\sexpy[(\hat{\tau} - \tau)^2 | \W] | \W \in \zb \backslash \zbo] \\
		 &\geq p \sexp_{\zbo}[\sexpy[(\hat{\tau} - \tau)^2 | \W]] + (1-p) q_\alpha \\
		 &\geq \sexp_{\zbo}[\sexpy[(\hat{\tau} - \tau)^2 | \W]] \\
		 &= \sexpy[\svar_{\zbo}[\hat{\tau} | \Y]]
\end{flalign*}

which concludes the proof.

\end{proof}

\subsection{Proof of Lemma~\ref{lemma:trick}}

\noindent\textbf{Statement of Lemma~\ref{lemma:trick}}

\bigskip

For $\W$ in $\mathcal{Z}^b$, we have: ~$\hat{\tau}(1-\W) = 2\cdot\tau - \hat{\tau}(\W)$.

\bigskip

\begin{proof}

Let $\W \in \mathcal{Z}^b$, and let $\W^* = 1 - \W$. Clearly, we have $N_1(1-\W) = N_1(\W) = \frac{N}{2}$, and $N_0(\W) = N_0(1-\W) = \frac{N}{2}$. And so:
\begin{flalign*}
	\hat{\tau}(\W) + \hat{\tau}(1-\W) &= \left(\frac{1}{N/2} \sum_{Z_i=1} Y_i(1) - \frac{1}{N/2}\sum_{Z_i=0}Y_i(0) \right) + \left(\frac{1}{N/2} \sum_{1-Z_i=1} Y_i(1) - \frac{1}{N/2}\sum_{1-Z_i=0}Y_i(0) \right)\\
	&=  \left(\frac{1}{N/2} \sum_{Z_i=1} Y_i(1) - \frac{1}{N/2}\sum_{Z_i=0}Y_i(0) \right) + \left(\frac{1}{N/2} \sum_{Z_i=0} Y_i(1) - \frac{1}{N/2}\sum_{Z_i=1}Y_i(0) \right) \\
	&= \frac{1}{N/2} \sum_i^N Y_i(1) - \frac{1}{N/2} \sum_i^N Y_i(0) \\
	&= 2 \tau
\end{flalign*}
which completes the proof.
\end{proof}


\subsection{Proof of Theorem~\ref{th:theory}}
\label{appendix:proof-thm}
There are four parts to this theorem: we must show that the estimator is unbiased under the uniform designs on $\zb$, $\zbu$, $\zbo$, and $\zbuo$. Th proofs follow the same general ideas, so we will skip the details 
whenever the proofs are similar.

\bigskip

\noindent\textbf{(i) Unbiasedness under the uniform distribution on $\zb$}

\bigskip

\begin{proof}

This proof could be carried exactly as above. The longer proof that we use introduces concepts that will be reused in most of the following proofs, but in a simple scenario. \\
By definition we have:
\begin{equation}
	\W \in \mathcal{Z}^b \Rightarrow \W^* = 1 - \W \in \mathcal{Z}^b
\end{equation}
Now, introduce for all i the sets:
\begin{equation}
 \mathcal{Z}_{i=1}^b = \{ \W / \,\, \W \in \mathcal{Z}^b \,\,\, \mbox{and} \,\,\, \W_i=1\} \,\,\, \mbox{and} \,\,\,  \mathcal{Z}_{i=0}^b = \{ \W / \,\, \W \in \mathcal{Z}^b \,\,\, \mbox{and} \,\,\, \W_i=0\}
\end{equation}

Notice that $\W \in \mathcal{Z}_{i=1}^b \Leftrightarrow \W^* \in \mathcal{Z}_{i=0}^b$, which implies $|\mathcal{Z}_{i=1}^b| = |\mathcal{Z}_{i=0}^b|$. And since we also have:

\begin{equation}
\mathcal{Z}_{i=1}^b \cup \mathcal{Z}_{i=0}^b = \mathcal{Z}^b \,\,\, \mbox{ and } \,\,\, \mathcal{Z}_{i=1}^b \cap \mathcal{Z}_{i=0}^b = \emptyset
 \end{equation}
 we conclude that:

\begin{equation}
	|\mathcal{Z}_{i=1}^b| = |\mathcal{Z}_{i=0}^b| = |\mathcal{Z}^b| / 2
\end{equation}

Which implies that

\begin{equation}
	P_{\zb}(Z_i = 1) =  \frac{|\mathcal{Z}_{i=1}^b| }{|\mathcal{Z}^b| } = \frac{1}{2} = \frac{|\mathcal{Z}_{i=0}^b| }{|\mathcal{Z}^b| } = P_{\zb}(Z_i = 0) 
\end{equation}

for all i. We can then write:

\begin{flalign*}
	\sexp_{\zb}[\hat{\tau}] &= \sexp_{\zb}\left[ \frac{2}{N} \sum_{i=1}^N \left(I(Z_i=1)Y_i(1) - I(Z_i=0) Y_i(0) \right) \right]\\
	&= \frac{2}{N} \sum_{i=1}^N \left( P_\zb(Z_i=1) Y_i(1) - P_\zb(Z_i=0) Y_i(0) \right)\\
	&= \frac{1}{N} \sum_{i=1}^N \left( Y_i(1) - Y_i(0) \right)\\
	&= \tau
\end{flalign*}

which completes the proof.
\end{proof}

\bigskip

\noindent\textbf{(iii) Unbiasedness under the uniform distribution on $\zbo$}

\bigskip

\begin{proof}
It is clear from the previous proof, that the key element of the proof is to show that:

\begin{equation}
	P_{\zbo}(Z_i = 1) = P_{\zbo}(Z_i = 0) = \frac{1}{2}
\end{equation}

for all i. For this, we by start proving that:

\begin{equation}
	\W \in \zbo \Rightarrow \W^* = 1 - \W \in \zbo
\end{equation}

%

By the Lemma, we have $\hat{\tau}(\W^*) = 2\tau - \hat{\tau}(\W)$. Now, let $\W\in \zbo$, we then have:
\begin{flalign*}
	\mbox{MSE}(\hat{\tau} | \W^*) &= \sexpy\left[(\hat{\tau}(\W^*) - \tau)^2 | \W\right]\\
	&= \sexpy\left[ (2\tau - \hat{\tau}(\W) - \tau)^2 | \W\right]\\
	&= \sexpy\left[ (\tau - \hat{\tau}(\W) )^2 | \W\right]\\
	&= \mbox{MSE}(\hat{\tau}| \W) \\
	&\leq q^{MSE}_{\alpha}
\end{flalign*}
which means that $\W^* \in \zbo$. So we have proved that:
\begin{equation}
	\W \in \zbo \Rightarrow \W^* = 1 - \W \in \zbo
\end{equation}

The rest of the proof unfolds exactly as in the proof of (i).
\end{proof}

\bigskip

\noindent\textbf{(ii) Unbiasedness under the uniform distribution on $\zbu$}

\bigskip

\begin{proof}
Here again, the key is to show that:

\begin{equation}
	P_{\zbu}(Z_i = 1) = P_{\zbu}(Z_i = 0) = \frac{1}{2}
\end{equation}
Let  $\W \in \zbu$, then by definition, we have:
\begin{equation}
	\mbox{Bias}(\hat{\tau}, \tau | \W) = \sexp[\hat{\tau}(\W) - \tau  | \W] = 0
\end{equation}
but then, since $\zbu \subset \zb$, by the Lemma, we have:
\begin{flalign*}
	\mbox{Bias}(\hat{\tau} | \W^*) &= \sexpy[\hat{\tau}(\W^*) - \tau | \W] \\
	&= \sexpy[2\tau - \hat{\tau}(\W) - \tau | \W] \\
	&= -\sexpy[\hat{\tau}(\W) - \tau] \\
	&= - \mbox{Bias}(\hat{\tau} | \W) \\
	&= 0
\end{flalign*}

which implies, that $\W^* \in \zbu$. So we have proved that:

\begin{equation}
	\W \in \zbu \Rightarrow \W^* = 1 - \W \in \zbu
\end{equation}

The rest of the proof follows as in the previous two proofs.

\end{proof}

\bigskip

\noindent\textbf{(iv) Unbiasedness under the uniform distribution on $\zbuo$}

\bigskip

This proof is exactly the same as the previous one. We will just prove the fact that if $\zbu \neq \emptyset$, then $\zbuo$ contains at least two elements. The proof is simple: \\

\noindent Let $\W \in \zbu$. There exists a $\W_0$ that minimizes the mean square error on the set $\zbu$ and so $\W_0 \in \zbuo$. But we have shown in the proof of (i) that for $\W \in \zb$, we have $\mbox{MSE}(\hat{\tau}, \tau | \W) = \mbox{MSE}(\hat{\tau} | \W^*)$. And so since $\W_0 \in \zbuo \subset \zb$, we have:

\begin{equation}
	\mbox{MSE}(\hat{\tau}| \W_0) = \mbox{MSE}(\hat{\tau}| \W_0^*) = \mbox{Min}_{\W\in \zbu}\mbox{MSE}(\hat{\tau} | Z)
\end{equation}
 
 which means that $\W^* \in \zbuo$, and so $|\zbuo| \geq 2$.

\end{document}